\documentclass[reprint,aps,prb,twocolumn,floats,amsmath,amssymb,superscriptaddress]{revtex4-2}
\usepackage[utf8]{inputenc}
\usepackage{graphicx}
\usepackage{epsfig}
\usepackage{amsfonts}
\usepackage{amsmath}
\usepackage{natbib}
\usepackage{multirow}
\usepackage{bm}
\usepackage{epstopdf}
\usepackage{ulem}
\usepackage{color}
\usepackage{hyperref} 

\definecolor{/green}{rgb}{0.0, 0.72, 0.92}

\begin{document}
\title{Tuning two-dimensional electron and hole gases at LaInO$_{3}$/BaSnO$_{3}$ interfaces by polar distortions, termination, and thickness}
\author{Wahib Aggoune}
\email{aggoune@physik.hu-berlin.de}
\affiliation{Institut f\"{u}r Physik and IRIS Adlershof, Humboldt-Universit\"{a}t zu Berlin, 12489 Berlin, Germany}

\author{Claudia Draxl}
\affiliation{Institut f\"{u}r Physik and IRIS Adlershof, Humboldt-Universit\"{a}t zu Berlin, 12489 Berlin, Germany}
\affiliation{European Theoretical Spectroscopy Facility (ETSF)}
\date{\today}
\begin{abstract}
Two-dimensional election gases (2DEG), arising due to quantum confinement at interfaces between transparent conducting oxides, have received tremendous attention in view of electronic applications. Here, we explore the potential of interfaces formed by two lattice-matched wide-gap oxides of emerging interest, {\it i.e.}, the polar, orthorhombic perovskite LaInO$_{3}$ and the non-polar, cubic perovskite BaSnO$_{3}$, employing first-principles approaches. We find that the polar discontinuity at the interface is mainly compensated by electronic relaxation through charge transfer from the LaInO$_{3}$ to the BaSnO$_{3}$ side. This leads to the formation of a 2DEG hosted by the highly-dispersive Sn-\textit{s}-derived conduction band and a 2D hole gas of O-\textit{p} character, strongly localized inside LaInO$_{3}$. We rationalize how polar distortions, termination, thickness, and dimensionality of the system (periodic or non-periodic) can be exploited in view of tailoring the 2DEG characteristics, and why this material is superior to the most studied prototype LaAlO$_{3}$/SrTiO$_{3}$.
\end{abstract}
\maketitle


\section*{INTRODUCTION}
Heterostructures of transparent conducting oxides (TCOs) have attracted the attention of researchers in view of both fundamental research as well as potential applications~\cite{Ohtomo+04n,Mannhart+10sc}. Among them, interfaces of the perovskite materials LaAlO$_{3}$ and SrTiO$_{3}$ are most studied prototypes due to the emergence of fascinating physical phenomena including superconductivity, ferromagnetism~\cite{Bert+11np}, and interfacial two-dimensional electron gases (2DEG)~\cite{Ohtomo+04n}. The latter is mainly a consequence of the electronic reconstruction to compensate the interfacial polar discontinuity induced by deposition of the polar material LaAlO$_{3}$ formed by alternatingly charged (LaO)$^{+}$ and (AlO$_{2}$)$^{-}$ planes on a neutral, TiO$_{2}$-terminated SrTiO$_{3}$ substrate. The 2DEG density confined within the SrTiO$_{3}$ side of the interface can reach 0.5 electrons ($e$) per $a^{2}$ ($a$ being the lattice parameter of SrTiO$_{3}$), corresponding to $\sim 3.3 \times 10^{14}\mathrm{cm^{-2}}$ for a complete compensation of the formal polarization induced within LaAlO$_{3}$~\cite{Nakagawa+06nm}. However, in real samples, the presence of defects impact both, polar discontinuity and electronic reconstruction, and thus carrier mobilities~\cite{Huijben+09am}. This includes cation intermixing~\cite{Chambers2+10ssr,Huijben+09am}, oxygen vacancies~\cite{Thiel+06sc}, edge dislocations~\cite{Thiel+09prl}, and changes in surface stoichiometry~\cite{Xie+13am}. Overall, the electron mobilities at such an interface are very sensitive to growth conditions~\cite{Huijben+09am}. Besides these extrinsic effects, the low interfacial mobility of this material system is also caused by the low-dispersion (large effective electron masses) of the partially occupied Ti-\textit{d} states. Further, scattering of electrons within these bands induced by significant electron-phonon coupling (EPC) decreases the mobility from $10^{4}~\mathrm{cm^{2}V^{-1}s^{-1}}$ at low temperature, to $1~\mathrm{cm^{2}V^{-1}s^{-1}}$ at room temperature~\cite{Ohtomo+04n,Cancellieri+16nc}. 

According to the polar-catastrophe model, in a perfect LaAlO$_{3}$/SrTiO$_{3}$ interface, the formal polarization ($P^{0}_{\mathrm{LAO}}$) allows for an insulator-to-metal transition at the interface beyond a certain thickness ($t_{\mathrm{c}}$) of LaAlO$_{3}$. Estimating this quantity by considering the energy difference between the valence-band edge of LaAlO$_{3}$ and the conduction-band edge of SrTiO$_{3}$~\cite{Nakagawa+06nm,Schmitt+12nc}, one obtains a value of 3.5 unit cells. Higher $t_{\mathrm{c}}$ values reported theoretically and experimentally, are due to structural relaxations, {\it i.e.}, polar distortions that induce a polarization opposite to $P^{0}_{\mathrm{LAO}}$. Such polar distortions that are not considered in the polar-catastrophe model, maintain the insulating character of the interface above 3.5 unit cells as confirmed theoretically ~\cite{Pickett+09prl,Stengel+11prl} and later observed experimentally~\cite{Cantoni+12am}. These distortions appear mainly in the LaAlO$_{3}$ side of the interface and arise due to changes inter-plane distances between La and Al planes upon interface formation. More recently, a competition between polar and nonpolar distortions through octahedra tilts has been observed~\cite{Gazquez+17prl}. Interestingly, the dependence of the octahdra tilts on the LaAlO$_{3}$ thickness can be exploited to tune the functionality of such interface~\cite{Gazquez+17prl}. We expect fascinating characteristics to occur when involving a polar material with pristine octahedral distortions such as a perovskite with an orthorhombic primitive cell.

Focusing first on nonpolar candidates as the substrate, cubic BaSnO$_{3}$ (BSO) has emerged as a most attractive system to overcome the limitations of SrTiO$_{3}$, as it exhibits extraordinary room-temperature mobilities, reaching 320$~\mathrm{cm^{2}V^{-1}s^{-1}}$ ~\cite{Hkim+12ape,paudel+17prb,yaqin+16pccp,Krish+16apl,Useong+15apl,Lee+17armr}. This value is the highest ever measured in a TCO and attributed to the low effective electron mass as well as to the long relaxation time of the longitudinal optical phonon scattering compared to SrTiO$_{3}$~\cite{Hkim+12ape,Niedermeier+17aps,Niedermeier+16arxiv}. In contrast to other suggested polar materials to be combined with BaSnO$_{3}$, such as LaScO$_{3}$~\citep{paudel+17prb}, LaGaO$_{3}$~\cite{yaqin+16pccp}, or LaAlO$_{3}$~\cite{chambers+16apl}, LaInO$_{3}$ (LIO) has the advantage of being nearly lattice matched to BaSnO$_{3}$ and exhibiting a favorable band offset to confine a 2DEG within the BaSnO$_{3}$ side~\cite{Krish+16apl,Useong+15apl,Lee+17armr,kim+18apl,Markurt+19sr}. Interestingly, it has an orthorhombic structure with tilted InO$_{6}$ octahedra, thus ideal for exploring also interfaces made of tilted and nontilted components.

In a previous work~\cite{Martina+20prm}, it was shown by transmission electron microscopy (TEM) that LaInO$_{3}$ can be coherently grown on (001) BaSnO$_{3}$, forming a sharp interface with negligible atomic disorder or misfit dislocations. This characteristic makes such an interface fascinating, since --as reported for LaAlO$_{3}$/SrTiO$_{3}$~\cite{Thiel+09prl}-- the interface conductivity and the mobility of the electron gas are enhanced by minimizing the dislocation density. Later it was shown by HRTEM analysis~\cite{Martina+21} that even if the BaO termination of the (001) BaSnO$_{3}$ surface is most favorable, the LaInO$_{3}$/BaSnO$_{3}$ interface (termed LIO/BSO from now on) is characterized by the LaO/SnO$_{2}$ termination, which is key for the formation of a 2DEG. Thereby, cation intermixing at the interface was rated to be negligible. Therefore, the combination of LaInO$_{3}$ and BaSnO$_{3}$ exhibits all the key features and appears superior to LaAlO$_{3}$/SrTiO$_{3}$ interfaces for reaching a high mobility 2DEG. 

In this work, we explore the characteristics of ideal LIO/BSO interfaces, based on density-functional-theory (DFT), also employing many-body perturbation theory where needed. We focus only on intrinsic effects that may play a role in compensating the interfacial polar discontinuity, {\it i.e.}, electronic reconstruction (formation of 2DEG) and possible structural distortions (formation of a depolarization field). Considering first periodic heterostructures, we address the competition between the polar distortions and the 2DEG charge density to compensate the interfacial polar discontinuity, upon increasing the thickness of the polar LaInO$_{3}$ block. Second, we discuss the impact of the interface termination that may give rise to either a 2DEG or 2D hole gas (2DHG). Finally, motivated by the advancements in synthesis techniques and in control of nanoscale structures~\cite{Campbell+18nm}, we provide a detailed comparison between the characteristics of the 2DEG in a periodic heterostructure and a non-periodic LIO/BSO interface, discussing how one can exploit dimensionality for tailoring the properties of the 2DEG. Overall, our results demonstrate the potential of this material combination for tuning and achieving high electron density and mobility.
 
\section*{RESULTS AND DISCUSSION}
\textbf{Pristine materials}

Before discussing the results for the periodic heterostructures LIO/BSO, we summarize the basic properties of the pristine materials. BaSnO$_{3}$ crystallizes in the cubic space group Pm$\overline{3}$m and is built by alternating neutral (BaO)$^{0}$ and (SnO$_{2}$)$^{0}$ planes along the cartesian directions, making it a nonpolar material. Its static dielectric constant was estimated to be about 20~\cite{Hkim+12ape}. The calculated lattice constant of 4.119~\AA~obtained with the PBEsol functional is in excellent agreement with experiment~\cite{Hkim+12ape}. LaInO$_{3}$ has an orthorhombic perovskite structure of space group Pbnm, containing four formula units per unit cell. The optimized structural parameters \textit{a}= 5.70~\AA, \textit{b}= 5.94~\AA, and \textit{c}= 8.21~\AA\ are in good agreement with experimental values~\cite{Zbigniew+20jcg}. The corresponding pseudocubic unit cell is defined such to have the same volume per LaInO$_{3}$ formula as the orthorhombic structure [for more details, we refer to the Supporting Information (SI)~\cite{note-si}]. Its calculated lattice parameter is about 4.116~\AA. In LaInO$_{3}$, the InO$_{6}$ octahedra are tilted along the pseudocubic unit cell directions with an $a^{-}a^{-}c^{+}$ pattern according to the Glazer notation~\cite{glazer+72acsb}. Considering the formal ionic charges of La (+3), In (+3), and O (-2), the charged (LaO)$^{+1}$ and (InO$_{2}$)$^{-1}$ planes along the pseudocubic unit cell [100], [010], and [001] directions make LaInO$_{3}$ a polar material (see supplementary figure 1). 

The calculated lattice mismatch between the cubic BaSnO$_{3}$ and the pseudocubic LaInO$_{3}$ unit cells is about 0.07\%, suggesting a coherent interface as confirmed experimentally~\cite{Martina+20prm}. In the latter work, it was shown that LaInO$_{3}$ can be favorably grown on top of a (001) BaSnO$_{3}$ substrate along the three pseudocubic unit cell directions, preserving the polar discontinuity at the interface, regardless of the orientation. Here, we consider the interface formed by [001]-oriented LaInO$_{3}$ ({\it i.e.}, \textit{c}= 8.21~\AA\ corresponds to the out-of-plane direction) on top of the (001) BaSnO$_{3}$ surface (see supplementary figure 1). 

The electronic properties of both bulk BaSnO$_{3}$ and LaInO$_{3}$ have been reported by us in previous works~\cite{Aggoune+BSO,Aggoune+LIO}, revealing PBEsol band gaps of 0.9 and 2.87~eV, and quasiparticle band gaps of 3.5 and 5.0 eV, respectively, as obtained by the $G_0W_0$ correction to results obtained by the exchange-correlation functional HSE06 ($G_0W_0$@HSE06). In both cases, the valence-band maximum is dominated by O-\textit{p} states, while, the conduction-band minimum has Sn-\textit{s} and In-\textit{s} character, respectively. \\

\textbf{Stoichiometric periodic heterostructures}

\label{stoic-heterostructure.}

%
\begin{figure*}
 \begin{center}
  \includegraphics[width=.98\textwidth]{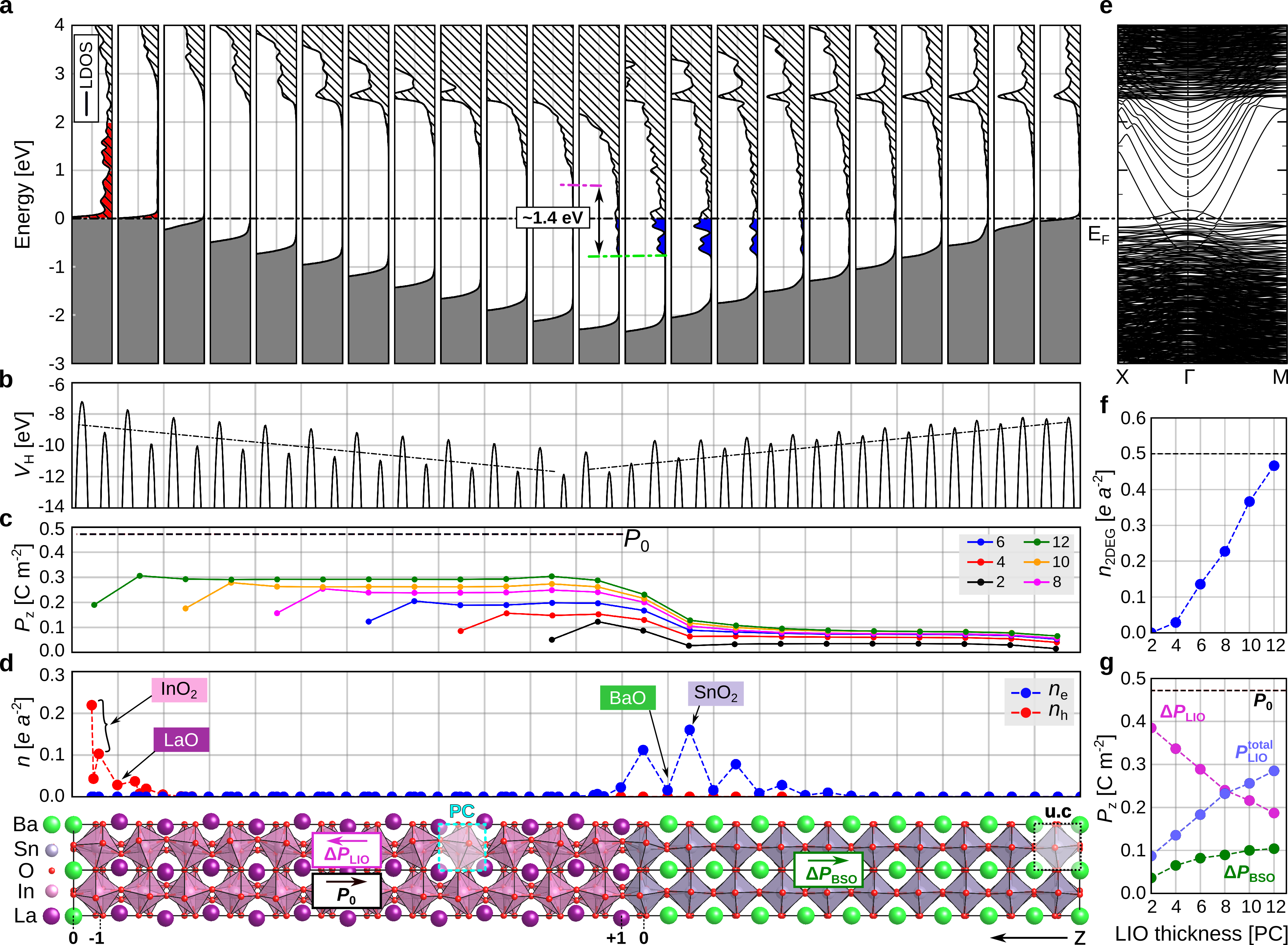}%
\caption{Stoichiometric periodic heterostructure: the system is formed by LIO$_{12}$/BSO$_{10}$ in out-of-plane direction \textit{z} (bottom), that is shared between panels (a), (b), (c), and (d) as their horizontal axis. $P_{0}$ is the formal polarization oriented from the (InO$_{2}$)$^{-1}$ to the (LaO)$^{+1}$ plane. (a) Local density of states per unit cell (LDOS) where the Fermi level is set to zero (also in panel (e)). The shaded gray area indicates the occupied valence band. The partially unoccupied valence (hole) and the occupied conduction (electron) states, resulting from electronic reconstruction, are highlighted by red and blue color, respectively. The dashed green and pink lines indicate the alignment of the conduction-band edges of BaSnO$_{3}$ and LaInO$_{3}$ at the interface. (b) In-plane averaged electrostatic potential along the \textit{z} direction. The dashed lines are guides to the eye, highlighting the trend of the potential. (c) Total polarization per unit cell computed for different LaInO$_{3}$ thicknesses (in units of the pseudocubic unit cell, PC). (d) Distribution of the electron (hole) charge densities obtained by integrating over the occupied conduction (unoccupied valence) states shown with blue (red) colored area in panel (a). (e) Electronic band structure along X-$\Gamma$-M. (f) Density of the 2DEG in electrons per BaSnO$_{3}$ surface unit cell ($e$ per $a^{2}$). (g) Total polarization ($P^{\mathrm{LIO}}_{\mathrm{total}}$, violet) within the LaInO$_{3}$ block and changes of polar distortions within the LaInO$_{3}$ ($\Delta P_{\mathrm{LIO}}$, magenta) and BaSnO$_{3}$ ($\Delta P_{\mathrm{BSO}}$, green) side as a function of the LaInO$_{3}$ thickness. The respective orientations are shown in the structural model.}
\label{fig:LaIn}
 \end{center}
\end{figure*}

In Fig.~\ref{fig:LaIn} we show a comprehensive compilation of the electronic properties of the LIO/BSO interface in a periodic heterostructure. Thereby, n-type (LaO/SnO$_{2}$) and p-type (InO$_{2}$/BaO) building blocks are periodically replicated in the out-of-plane direction (\textit{z}). In this case, the system is stoichiometric and ideally suited for investigating the electronic reconstruction due to the polar discontinuity. The BaSnO$_{3}$ block has a thickness of 10 unit cells which is enough to minimize the interaction with its replica. It is determined by making sure that the central part of the BaSnO$_{3}$ block behaves like in its bulk counterpart (see supplementary figure 2). The LaInO$_{3}$ block has a thickness of 12 pseudocubic unit cells. (We use the notation LIO$_{12}$/BSO$_{10}$ heterostructure in the following). Since the LaInO$_{3}$ block has different terminations on the two sides, a formal polarization of $P_{0}=e/2a^{2}$=0.47 C~m$^{-2}$, oriented from the (InO$_{2}$)$^{-1}$ to the (LaO)$^{+1}$ plane, is induced (black arrow in the bottom panel of Fig.~\ref{fig:LaIn}). As BaSnO$_{3}$ is a nonpolar material, this gives rise to a polar discontinuity at the interface.

The charge reconstruction is evident from the electronic band structure [Fig.~\ref{fig:LaIn}(e)] showing that the combination of these two insulators has metallic character. From the local density of states (LDOS, per unit cell) depicted in panel (a) along the \textit{z} direction, we can clearly see that the dipole induced within the LaInO$_{3}$ side causes an upward shift of the valence-band edge, evolving between the (LaO)$^{+1}$ and (InO$_{2}$)$^{-1}$ terminations. This is also reflected in the in-plane averaged electrostatic potential shown in panel (b) (for more details see supplementary figure 9). At the latter termination, the valence-band maximum crosses the Fermi level inside LaInO$_{3}$, leading to a charge transfer to the BaSnO$_{3}$ side in order to compensate the polar discontinuity. Consequently, the bottom of the conduction band of BaSnO$_{3}$ becomes partially occupied, giving rise to a 2DEG confined within three unit cells ($\sim$10~\AA). A 2DHG forms in the LaInO$_{3}$ side localized within one pseudocubic unit cell ($\sim$4~\AA). Integrating over these now partially occupied conduction states (see supplementary methods), we find that the 2DEG density reaches a value of $2.7\times10^{14}~\mathrm{cm^{-2}}$ \textit{i.e.}, $\sim$0.46~$e$ per $a^{2}$ ($a^2$ being the unit cell area of bulk BaSnO$_{3}$). Obviously, the same value is obtained for the 2DHG when integrating over the now empty parts of the valence bands. The charge distribution shown in Fig.~\ref{fig:LaIn}(d) reveals that the 2DEG is located mainly within the SnO$_{2}$ plane and exhibits Sn-\textit{s} character. This highly-dispersive $s$-band suggests high mobility, unlike the situation in LaAlO$_{3}$/SrTiO$_{3}$. These results demonstrate the exciting potential of such a material combination as an ideal platform for achieving a high-density 2DEG. We also highlight the importance of the well-confined 2DHG hosted by O-\textit{p} states in view of p-type conductivity. We note that in SrTiO$_{3}$/LaAlO$_{3}$/SrTiO$_{3}$, another heterostructures~\cite{Maznichenko+20pssb} formed by a polar and a nonpolar material, only recently the presence of a 2DHG has been confirmed experimentally~\cite{Campbell+18nm}. The coexistence of high-density well-confined electron and hole gases within one system as shown here, appears as a promising platform for exploring also intriguing phenomena such as long-lifetime bilayer excitons~\cite{Millis+10prb} or Bose–Einstein condensation~\cite{Eisenstein+04n}.

\textbf{Polar distortions in stoichiometric periodic heterostructures}

The 2DEG density reached in the periodic heterostructure LIO$_{12}$/BSO$_{10}$, being slightly lower than 0.5 $e$ per $a^{2}$, indicates that polar distortions are involved to partially compensate for the polar discontinuity. Looking into its optimized geometry, we find that the tilt of the octahedra decreases gradually from the LaInO$_{3}$ to the BaSnO$_{3}$ side [see Fig.~\ref{fig:LaIn}(bottom panel)]. Consequently, the out-of-plane lattice spacing increases close to the interface by about 3\% (see supplementary figure 2), in good agreement with an experimental observation~\cite{Markurt+19sr}. We also find that the out-of-plane displacements of the inequivalent O atoms are not the same within all octahedra (see supplementary figure 2 and 3). Moreover, the distances between AO and BO$_{2}$ planes (A= Ba, La and B= Sn, In) across the interface are also unequal (see supplementary figure 2). Using a simple linear approximation for the local polarization based on Born effective charges ($Z^{*}$)~\cite{berryPhase+93prbr,Resta+93prl} of the atomic species (calculated for the pristine materials), we obtain a qualitative trend of the out-of-plane polarization induced by such structural distortions (see supplementary discussion). We note that due to the tilt of the octahedra, calculating the polarization for such a heterostructure is less straightforward than for, {\it e.g.}, LaAlO$_{3}$/SrTiO$_{3}$. We find that the structural distortions within the LaInO$_{3}$ side induce a polarization $\Delta P_{\mathrm{LIO}}$ that counteracts the formal polarization $P_{0}$. Moreover, the polar discontinuity at the interface is reduced by structural distortions within the BaSnO$_{3}$ side, inducing $\Delta P_{\mathrm{BSO}}$ that is parallel to $P_{0}$. The total polarization within the LaInO$_{3}$ side, $P^{\mathrm{LIO}}_{\mathrm{total}}$, shown in Fig.~\ref{fig:LaIn}(g) is the sum of $-\Delta P_{\mathrm{LIO}}$ and $P_{0}$. As expected for the particular case of LIO$_{12}$/BSO$_{10}$, the average polarization within LaInO$_{3}$ ($P^{\mathrm{LIO}}_{\mathrm{total}}$) is smaller than $P_{0}$ due to partial compensation by structural distortions. For this reason, the 2DEG density mentioned above is smaller than 0.5 $e$ per $a^{2}$. For better grasping the polar discontinuity at the interface, we plot the total polarization along the \textit{z} direction [see Fig.~\ref{fig:LaIn}(c)]. Focusing first on the particular case of LIO$_{12}$/BSO$_{10}$, we can clearly see that within the LaInO$_{3}$ side, it is smaller than $P_{0}$ and also non-negligible inside BaSnO$_{3}$, making the polar discontinuity at the interface less pronounced. As we provide only a qualitative analysis of the polarization, we do not expect a full match between the values of the 2DEG and the corresponding polarization discontinuity. However, the obtained trend of the polarization strength is valuable to understand and explain the relationship between the calculated 2DEG density and the LaInO$_{3}$ thickness.\newline

\textbf{Impact of the LaInO$_{3}$ thickness in stoichiometric periodic heterostructures}
\label{thickness}

Now, we fix the thickness of the BaSnO$_{3}$ building block to 10 unit cells and vary that of LaInO$_{3}$ between 2 and 12 pseudocubic unit cells, labeling the systems LIO$_{m}$/BSO$_{10}$ ($m$=2, 4, 6, 8, 10, 12). Before discussing the results, we note that the \textit{critical} thickness, $t_{\mathrm{c}}$, for an insulator-to-metal transition at the interface is about one pseudocubic LaInO$_{3}$ unit cell, when ignoring effects from structural relaxation as given by the polar catastrophe model~\cite{Nakagawa+06nm,Schmitt+12nc}. This value is obtained as $t_{\mathrm{c}}=\epsilon_{0}\epsilon_{\mathrm{LIO}}\Delta E/eP^{0}_{\mathrm{LIO}}$. Here, $\epsilon_{\mathrm{LIO}}\sim24$ is the relative static dielectric constant of LaInO$_{3}$~\cite{jang+17jap,Zbigniew+20jcg}, and $\Delta E$= 1 eV represents the energy difference between the valence-band edge of LaInO$_{3}$ and the conduction-band edge of BaSnO$_{3}$ at the interface, obtained by the PBEsol functional [see Fig.~\ref{fig:LaIn}(a)]. In contrast to the \textit{unrelaxed} LIO$_{2}$/BSO$_{10}$ system, where we find metallic character, with full electronic reconstruction of 0.5 $e$ per $a^{2}$ (see supplementary figure 4), atomic relaxations lead to semiconducting character, {\it i.e.}, no formation of a 2DEG [see Fig.~\ref{fig:LaIn}(f)]. We observe that the polar distortions that counteract the formal polarization $P_{0}$, hamper the electronic reconstruction at the interface and, thus, the formation of a 2DEG, up to a \textit{critical} thickness of 4 pseudocubic LaInO$_{3}$ unit cells [see Fig.~\ref{fig:LaIn}(f)]. This result highlights the importance of structural relaxations for compensating the polar discontinuity and stabilizing the interface. 

Focusing on the electronic charge, we find that at a thickness of 4 pseudocubic LaInO$_{3}$ unit cells, the density of the 2DEG is only about 0.03 $e$ per $a^{2}$ [see Fig.~\ref{fig:LaIn}(f)], {\it i.e.}, distinctively smaller than the nominal value of 0.5 $e$ per $a^{2}$. Increasing the LaInO$_{3}$ thickness, the 2DEG density increases progressively and reaches a value of $\sim$0.46 $e$ per $a^{2}$ with 12 pseudocubic LaInO$_{3}$ unit cells. This means that the polar distortions are also non-negligible beyond the \textit{critical} thickness. In Fig~\ref{fig:LaIn}(g), we display the averages of $\Delta P_{\mathrm{LIO}}$ and $\Delta P_{\mathrm{total}}$ for the considered structures LIO$_{m}$/BSO$_{10}$, finding that the polar distortions (total polarization) is maximal (minimal) at two pseudocubic LaInO$_{3}$ unit cells and decreases (increases) with LaInO$_{3}$ thickness. $\Delta P_{\mathrm{BSO}}$ increases with LaInO$_{3}$ thickness, but it is smaller than its counterpart in LaInO$_{3}$. As a result, the polar discontinuity at the interface increases with increasing LaInO$_{3}$ thickness [see Fig.~\ref{fig:LaIn}(c)]. Hence, the 2DEG density increases accordingly in order to compensate it [see Fig.~\ref{fig:LaIn}(f)].

For a more reliable estimation of the \textit{critical} thickness of LaInO$_{3}$ for an insulator-to-metal transition, we evaluate $\Delta E$ by considering the quasiparticle band gaps of the constituents~\cite{Aggoune+BSO,Aggoune+LIO}. Applying a scissor shift to the PBEsol band offset leads to a quasiparticle value of $\Delta E$=3.4 eV, in agreement with the band offset reported in Ref.~\cite{Useong+15apl} (see supplementary discussion). Thus, we obtain a $t_{\mathrm{c}}$=4 pseudocubic LaInO$_{3}$ unit cell which is increased by 3 pseudocubic unit cells compared to that given by PBEsol offset (one pseudocubic unit cell). For the relaxed structures, we estimate $t_{\mathrm{c}}$ to be about seven pseudocubic LaInO$_{3}$ unit cells when considering quasiparticle band gaps, which is distinctively larger than the 4 pseudocubic unit cells derived from PBEsol [see Fig.~\ref{fig:LaIn}(f)]. This result indicates that a thick LaInO$_{3}$ component is needed to reach high 2DEG densities in periodic heterostructures, as both sides contribute to the compensation of the polar discontinuity through atomic distortions. This result is inline with a previous theoretical discussion reported for oxide interfaces~\cite{Stengel+11prl}.

Proceeding now to the nature of the atomic distortions, we find that, in contrast to the LaAlO$_{3}$/SrTiO$_{3}$ interface where the unequal distances between La and Al planes dominate~\cite{Cantoni+12am}, the unequal displacements of the inequivalent oxygen atoms are most decisive for the polar distortions in the LaInO$_{3}$ side of the interface (see supplementary figure 2). This behavior is governed by the gradual tilts of octahedra across the interface which facilitates the compensation of the polar discontinuity. This indicates that below the \textit{critical} thickness, this compensation happens through such atomic distortions, rather than elimination by ionic intermixing or other defects, explaining the sharp interface and negligible intermixing observed experimentally~\cite{Martina+20prm}. The latter characteristic is crucial for achieving a high-density 2DEG beyond $t_{\mathrm{c}}$. The band offset at the interface shows that the conduction-band minimum of BaSnO$_{3}$ is about 1.4 eV below that of LaInO$_{3}$, confining the 2DEG at the BaSnO$_{3}$ side [see Fig.~\ref{fig:LaIn}(a)]. 
\begin{figure*}
 \begin{center}
 \includegraphics[width=.98\textwidth]{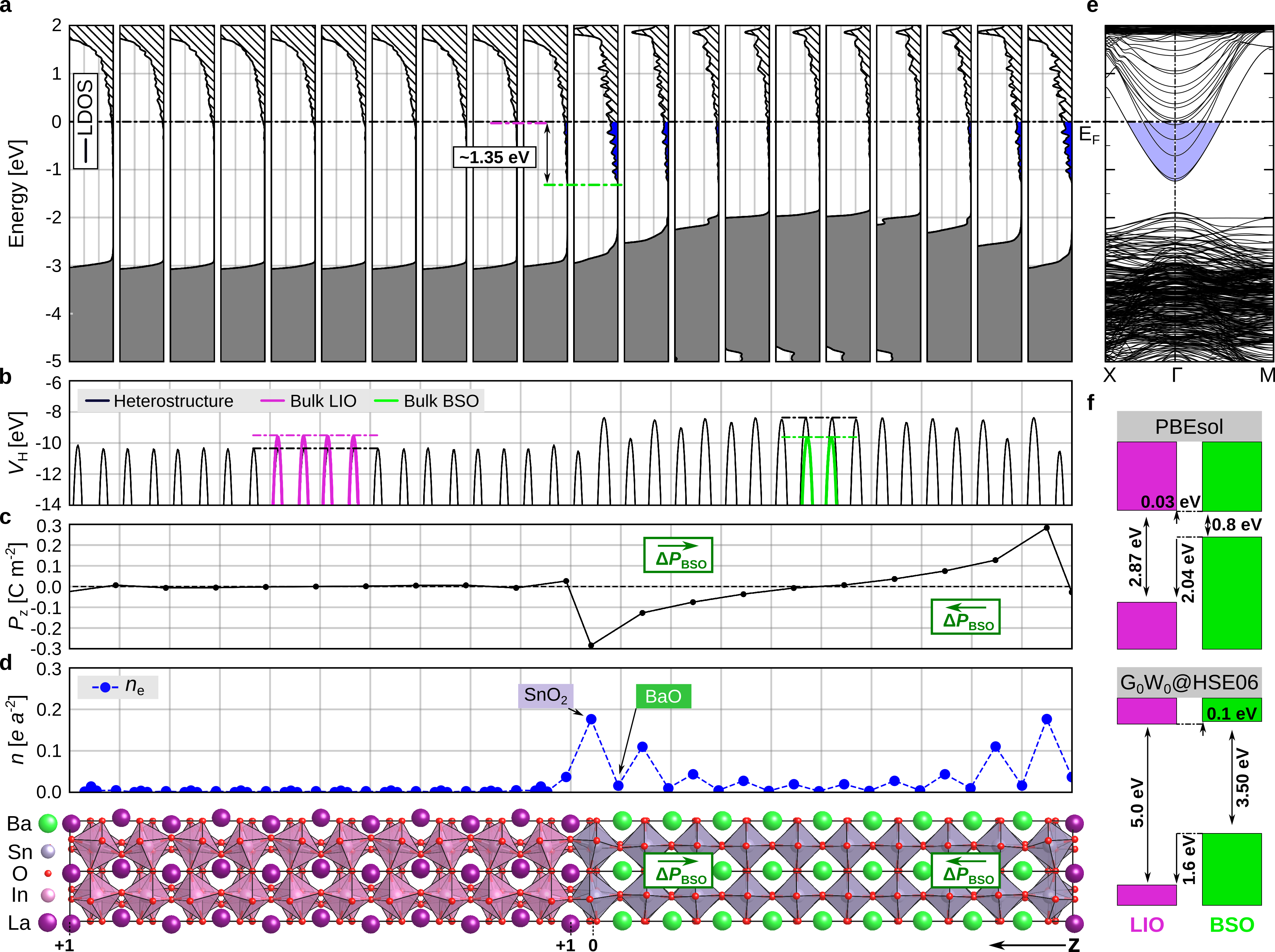}%
\caption{Non-stoichiometric periodic nn-type heterostructure: the system is formed by LIO$_{10}$/BSO$_{10}$ in out-of-plane direction \textit{z} (bottom), that is shared between panels (a), (b), (c), and (d) as their horizontal axis. The LaInO$_{3}$ block is terminated by a (LaO)$^{+1}$ plane on both sides. (a) Local density of states per unit cell (LDOS). The Fermi level is set to zero, and the electron population of the (original) conduction bands is shown as shaded blue area (also in panel (e)). The dashed green and pink lines indicate the alignment of the conduction-band edges of the two materials at the interface. (b) In-plane averaged electrostatic potential along the \textit{z} direction. (c) Total polarization per unit cell. (d) Distribution of the electron charge density obtained by integrating over the occupied conduction states indicated by the blue area in panel (a). (e) Electronic band structure along X-$\Gamma$-M. (f) Band alignment as obtained by using the band gaps of the bulk systems (PBEsol [top] and G$_{0}$W$_{0}$@HSE06 [bottom]), considering the energy difference between the potential of the bulk and the periodic heterostructure as shown in panel (b). The alignment at the interface obtained by PBEsol can be seen in panel (a).}
\label{fig:LaLa}
 \end{center}
\end{figure*}

\textbf{Non-stoichiometric nn-type periodic heterostructure}

Adopting thicker polar building blocks, {\it i.e.}, above 12 pseudocubic unit cells, the interaction between the n-type and the p-type interfaces is prevented. Due to the considerable computational costs, several models were proposed to predict the characteristics of such situation in oxide interfaces~\cite{Stengel+11prl}. One of them is to consider non-stoichiometric structures, where the polar LaInO$_{3}$ block is terminated by a (LaO)$^{+1}$ plane on both sides. In this case, termed nn-type periodic heterostructure, the system is self-doped as the additional (LaO)$^{+1}$ layer serves donor. As the LaInO$_{3}$ building block is symmetric, the formal polarization $P_{\mathrm{0}}$ is induced from the middle of the slab outwards on both sides, that compensate each other. In this way, the built-in potential inside the LaInO$_{3}$ is avoided, while the discontinuity at the LIO/BSO interface is preserved. To this end, we consider a periodic heterostructure formed by 10 BaSnO$_{3}$ unit cells and 10 pseudocubic LaInO$_{3}$ unit cells, which is large enough to minimize the interaction between the periodic n-type interfaces [see Fig.~\ref{fig:LaLa}(bottom) and supplementary figure 6]. 

The electronic band structure, obtained by PBEsol, shows that this system has metallic character, where the partial occupation of the conduction band amounts to 1 $e$ per $a^{2}$ [see Fig.~\ref{fig:LaLa}(e)]. The corresponding effective electron mass, being 0.24 m$_{e}$, is quite low compared with that of LaAlO$_{3}$/SrTiO$_{3}$ interfaces (0.38 m$_{e}$~\cite{Maznichenko+20pssb}), and suggests a high electron mobility. This value is close to that found for pristine BaSnO$_{3}$ (0.17 m$_{e}$ obtained by PBEsol, 0.2 m$_{e}$ by $G_{0}W_{0}$)~\cite{Aggoune+BSO}. In pristine SrTiO$_{3}$, a transition from band-like conduction (scattering of renormalized quasiparticles) to a regime governed by incoherent contributions due to the interaction between the electrons and their phonon cloud has been reported upon increasing temperature~\cite{Zhou+19prr}. In BaSnO$_{3}$, the relaxation time for the longitudinal optical phonon scattering is found to be larger compared to SrTiO$_{3}$, contributing to the high-room temperature mobility reported for the La-doped BaSnO$_{3}$ single crystals~\cite{Niedermeier+16arxiv}. Based on this, a high room-temperature mobility is also expected at the here investigated interfaces as this material combination basically preserves the structure of the pristine BaSnO$_{3}$~\cite{krish+17prb}. Overall, significant polaronic effects are not expected. In both constituents, we find typical EPC effects on the electronic properties~\cite{Aggoune+BSO,Aggoune+LIO}, {\it i.e.}, a moderate renormalization of the band gap by zero point vibrations and  temperature. Given the excellent agreement between theory and experiment that can explain all features of the optical spectra~\cite{Aggoune+BSO,Aggoune+LIO}, we conclude that polaronic distortions do not play a significant role. Thus, we do not expect a dramatically different behavior at the interfaces.

Before analyzing the spatial charge distribution, we note that in a pristine symmetric LaInO$_{3}$ slab, the electronic charge accumulates on its surfaces, accompanied by structural distortions that tend to screen the discontinuity of the polarization. Combined with BaSnO$_{3}$, the polar distortions vanish at the LaInO$_{3}$ side but appear in the BaSnO$_{3}$ building block and are accompanied by a charge redistribution [see Fig.~\ref{fig:LaLa}(a) and (c)]. In Fig.~\ref{fig:LaLa}(c), we display the polarization induced by the structural distortions along the slab. We find that the gradual decrease of the octahedra tilt from the LaInO$_{3}$ to the BaSnO$_{3}$ side enlarges the out-of-plane lattice spacing at the interface and enhances the unequal oxygen-cation displacements within the BaSnO$_{3}$ side (see supplementary figure 6). This, in turn, induces a gradually decreasing polarization ($\Delta P_{\mathrm{BSO}}$) oriented away from the interface. Due to this local dipole, the BaSnO$_{3}$ conduction-band edge is shifted downwards near the interface, leading to partial charge redistribution [see Fig.~\ref{fig:LaLa}(a) and (d)], and being also evident in the in-plane averaged electrostatic potential [panel (b)]. Integrating the LDOS over the partially occupied conduction states, we find that the 2DEG is confined within three unit cells, amounting to 0.5 $e$ per $a^{2}$ at each side [see Fig.~\ref{fig:LaLa}(d)]. Thereby the 2DEG is mainly hosted by Sn-\textit{s} states within the SnO$_{2}$ planes. The valence and conduction-band edges at the LaInO$_{3}$ side are almost the same in all pseudocubic unit cells as the polar distortions are negligible (only nonpolar tilts) [see Figs.~\ref{fig:LaLa}(a) and (c)]. These results are inline with the trend of the polarization and the 2DEG density obtained for the stoichiometric periodic system. While in the latter, we observed decreasing (increasing) structural distortions in the LaInO$_{3}$ (BaSnO$_{3}$) side as a function of LaInO$_{3}$ thickness [see Fig.~\ref{fig:LaIn}(g)], here, the polar distortions vanish (are enhanced) at the LaInO$_{3}$ (BaSnO$_{3}$) side of the interface. Consequently, the conduction-band minimum of the BaSnO$_{3}$ side is lowered by about 1.35 eV with respect to that of LaInO$_{3}$ [see Fig.~\ref{fig:LaLa}(a)]. This value is also inline with that estimated for the stoichiometric periodic heterostructure LIO$_{12}$/BSO$_{10}$ ($\sim$ 1.4 eV). Such an offset is crucial for confining the 2DEG in order to reach a value of 0.5 $e$ per $a^{2}$ [see Fig.~\ref{fig:LaLa}(a) and (d)].

Note that here, we can determine the band offsets by considering the respective quasiparticle gaps of the pristine materials~\cite{Aggoune+BSO,Aggoune+LIO} as well as an alternative approach based on the electrostatic potential~\cite{dima+19cm} (see Fig.~\ref{fig:LaLa}(b) and supplementary discussion). We find that, the conduction-band offset between the middle of the BaSnO$_{3}$ and LaInO$_{3}$ blocks is almost the same when using quasiparticle or PBEsol band gaps [see Fig.~\ref{fig:LaLa}(b) and (f)]. Thus, we conclude that PBEsol is good enough to capture band offset and charge distribution at the interface. The latter conclusions are confirmed by calculations using HSE06 for a smaller (feasible) nn-type system (see supplementary figure 7).

\textbf{Non-stoichiometric pp-type periodic heterostructure}

Calculations for a periodic pp-type heterostructure (see supplementary figure 5) with (InO$_{2}$)$^{-1}$ termination on both sides of LaInO$_{3}$ reveal a 2DHG with a density of 0.5 $e$ per $a^{2}$. Interestingly, the hole stays at the LaInO$_{3}$ side, confined within one pseudocubic unit cell. Like in the periodic nn-type heterostructure, the tilt of the octahedra decreases gradually from the LaInO$_{3}$ to BaSnO$_{3}$ side. However, the BaO termination in the pp-type case favors nonpolar distortions within the BaSnO$_{3}$ side, \textit{i.e.}, equal displacements of the inequivalent O atoms, while polar distortions appear only in the LaInO$_{3}$ side. Consequently, local dipoles are induced in the latter, pushing up its valence-band edge above that of BaSnO$_{3}$ (see supplementary figure 5). Hence, the 2DHG exhibiting O-\textit{p} character, stays on the LaInO$_{3}$ side of the interface. It has been shown recently by a combined theoretical and experimental investigation~\cite{Martina+21} that the n-type interface is more favorable than the p-type interface. Since at the BaO-terminated p-type interface, the 2DHG stays within the LaInO$_{3}$ side it contributes less to the compensation of the interfacial polar discontinuity. In contrast, the SnO$_2$-terminated n-type interface allows for electronic charge transfer to the BaSnO$_{3}$ side, forming a 2DEG that compensates the interfacial polar discontinuity more efficiently (see supplementary discussion).\newline

\textbf{Stoichiometric non-periodic interface}

\begin{figure*}
 \begin{center}
\label{fig:LaIn-IF}
 \includegraphics[width=.98\textwidth]{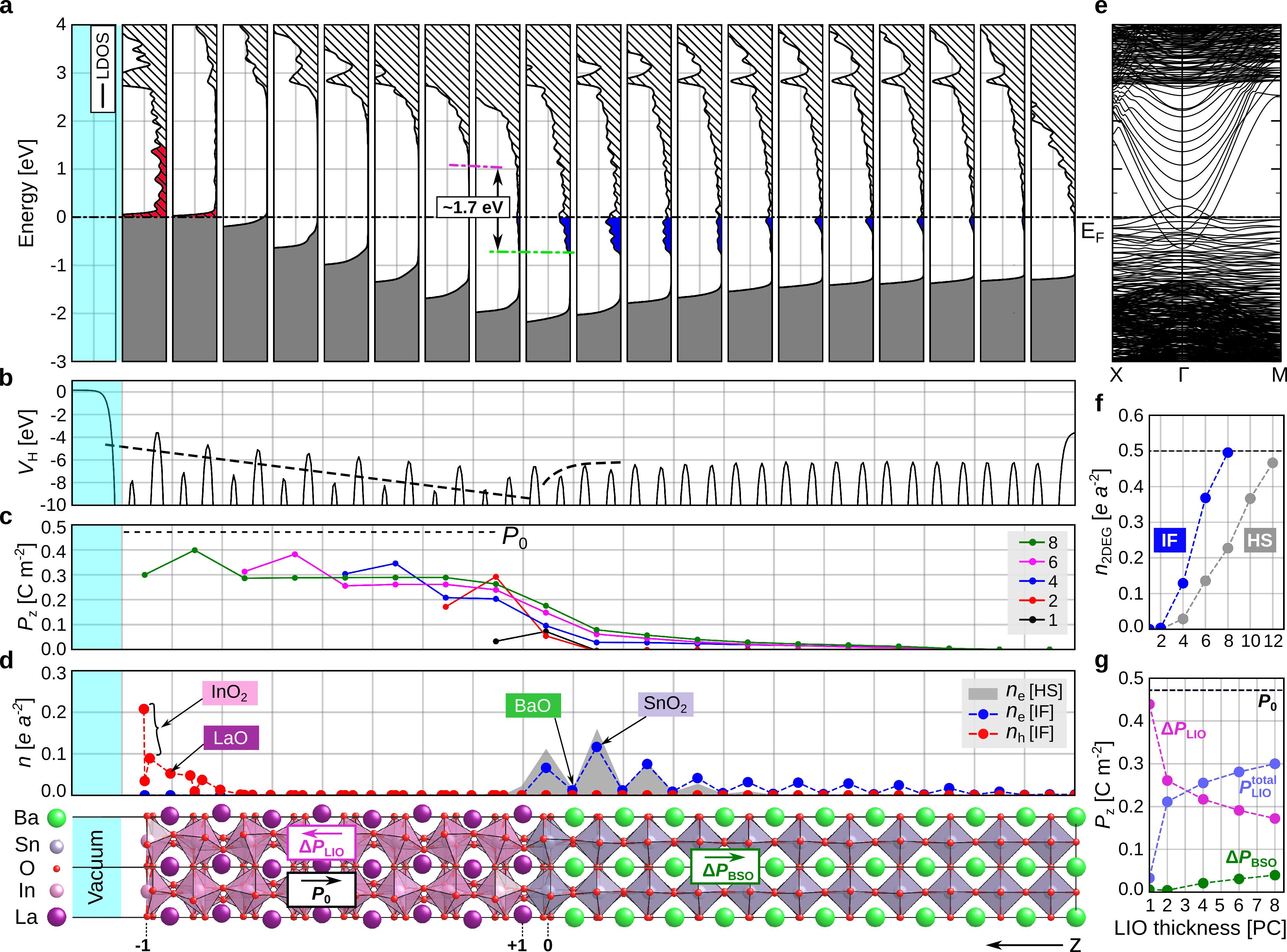}%
\caption{Stoichiometric non-periodic interface (here termed IF): the system is formed by LIO$_{8}$/BSO$_{11}$ in out-of-plane direction \textit{z} (bottom), that is shared between panels (a), (b), (c), and (d) as their horizontal axis. $P_{0}$ is the formal polarization oriented from the (InO$_{2}$)$^{-1}$ plane towards the (LaO)$^{+1}$ plane at the interface. (a) Local density of states per unit cell (LDOS) with the Fermi level being set to zero (also in panel (e)). The shaded gray area indicates the occupied valence states. The depleted valence band region (holes) and the occupied conduction band region (electrons) resulting from electronic reconstruction, are highlighted by red and blue color, respectively. The dashed green and pink lines indicate the alignment of the conduction-band edges of the two materials at the interface. (b) In-plane averaged electrostatic potential along the \textit{z} direction. (c) Total polarization per unit cell computed for different LaInO$_{3}$ thicknesses (in units of the pseudocubic unit cell, PC). (d) Distribution of the electron (hole) charge densities obtained by integrating the LDOS indicated by blue (red) color in panel (a). For comparison, the 2DEG distribution in the periodic heterostructure LIO$_{12}$/BSO$_{10}$ (here termed HS) is shown by the shaded gray area. (e) Electronic band structure along X-$\Gamma$-M. (f) Density of the 2DEG in $e$ per $a^{2}$ in the non-periodic (blue) and periodic (gray) heterostructures. (g) Total polarization ($P^{\mathrm{LIO}}_{\mathrm{total}}$, violet) within the LaInO$_{3}$ side and changes of polar distortions within LaInO$_{3}$ ($\Delta P_{\mathrm{LIO}}$, magenta) and BaSnO$_{3}$ ($\Delta P_{\mathrm{BSO}}$, green) side as a function of LaInO$_{3}$ thickness. The respective orientations are shown in the structural model.}
 \end{center}
\end{figure*}

Now, we investigate the case of a non-periodic LIO/BSO interface, consisting of a thin LaInO$_{3}$ layer on top of a (001) BaSnO$_{3}$ substrate. Considering stoichiometric systems, we only focus on the n-type interface as it is predicted to be more favorable~\cite{Martina+21}. For BaSnO$_{3}$, we find that 11 unit cells are enough to capture the extension of the structural deformations and the 2DEG distribution away from the interface. We then vary the thickness of LaInO$_{3}$ between one and eight pseudocubic unit cells, labeling the systems as LIO$_{n}$/BSO$_{11}$ ($n$=1, 2, 4, 6, 8). In Fig.~\ref{fig:LaIn-IF} (bottom panel), we show the optimized geometry of LIO$_{8}$/BSO$_{11}$. The BaSnO$_{3}$ substrate terminates with a (SnO$_{2}$)$^{0}$ plane at the interface, and the surface termination of LaInO$_{3}$ is a (InO$_{2}$)$^{-1}$ plane. Thus, LaInO$_{3}$ is stoichiometric and has a formal polarization of $P_{0}$=0.47~C~m$^{-2}$, oriented from the surface to the interface, as in the stoichiometric periodic heterostructures discussed above.

Electronic reconstruction, that leads to a metallic character, is evident from the resulting band structure, where the valence and conduction bands overlap at the $\Gamma$ point [see Fig.~\ref{fig:LaIn-IF}(e)]. In the LDOS, we clearly see that the dipole induced within LaInO$_{3}$ causes an upward shift of the valence-band edge (most pronounced at the surface) which is also evident in the in-plane averaged electrostatic potential [see Figs.~\ref{fig:LaIn-IF}(a) and (b)]. At the surface, the valence-band edge crosses the Fermi level, leading to a charge transfer across the interface that counteracts the polar discontinuity [see Fig.~\ref{fig:LaIn-IF}(a)]. Consequently, the bottom of the conduction band becomes partially occupied within the BaSnO$_{3}$ side, giving rise to a 2DEG that is confined within five unit cells ($\sim20$\:\AA) [see Fig.~\ref{fig:LaIn-IF}(d)]. The 2DHG formed at the surface is localized within one pseudocubic LaInO$_{3}$ unit cell ($\sim$4\:\AA). In this case, the conduction-band minimum of the BaSnO$_{3}$ building block is about 1.7 eV below that of LaInO$_{3}$ [Fig.~\ref{fig:LaIn}(a)], in agreement with an experimental observation of 1.6 eV~\cite{Useong+15apl}. Integrating the LDOS over the region of these partially filled states, we find that the 2DEG density (and likewise the 2DHG density) reaches a value of $2.9\times10^{14}~\mathrm{cm^{-2}}$ {\it i.e.}, $\sim$0.49 $e$ per $a^{2}$. 

By increasing the LaInO$_{3}$ thickness from one to eight pseudocubic unit cells, the polar distortions ($\Delta P_{\mathrm{LIO}}$) decrease, being accompanied by an electronic charge transfer from the surface to the interface [see Fig.~\ref{fig:LaIn-IF}(f) and (g)]. Compared to the periodic heterostructure [Fig.~\ref{fig:LaIn}(c)], they are less pronounced at the BaSnO$_{3}$ side [Fig.~\ref{fig:LaIn-IF}(c)]. This enhances the polar discontinuity (see supplementary figure 8) and thus, the 2DEG density that reaches a higher value than in the periodic systems with similar LaInO$_{3}$ thickness [see Fig.~\ref{fig:LaIn-IF}(f)]. Focusing on the charge confinement, the conduction-band edge in the BaSnO$_{3}$ side is gradually shifted up when moving away from the interface as $\Delta P_{\mathrm{BSO}}$ is less pronounced. This allows an extension of the 2DEG up to five unit cells ($\sim20~\AA$) in the BaSnO$_{3}$ (substrate) compared to three unit cells in the periodic heterostructure case [see Fig.~\ref{fig:LaIn-IF}(d)]. The enhanced polar discontinuity reduces the \textit{critical thickness} to only two pseudocubic LaInO$_{3}$ unit cells compared to four found for the periodic heterostructure, when relying on PBEsol [see Fig.~\ref{fig:LaIn-IF}(f)]. Based on the quasiparticle band gaps, however, we estimate $t_{\mathrm{c}}$ to be five pseudocubic LaInO$_{3}$ unit cells in the non-periodic system compared to seven in the periodic case. Overall tuning of the 2DEG charge density through the LaInO$_{3}$ thickness is possible, and remarkably, also the type of heterostructure ({\it i.e.}, periodic or non-periodic) impacts its spacial distribution. Both aspects can be exploited to tune the characteristic of the 2DEG. 

In view of realistic applications, it is worth considering the results presented here in the context of existing experimental research on LIO/BSO interfaces. A main challenge here is the quality of the BaSnO$_{3}$ substrate~\cite{galazka+16jpcm}. Previous experimental works~\cite{Useong+15apl,Markurt+19sr} investigated a field effect transistor, formed by a LaInO$_{3}$ gate and a La-doped BaSnO$_{3}$ channel on a SrTiO$_{3}$ substrate. An enhancement of conductance with increasing LaInO$_{3}$ thickness was observed, but no indication of a \textit{critical} thickness for an insulator-to-metal transition at the interface. A maximal 2DEG density of only 3 $\times 10^{13}$ (0.05 $e$ per $a^{2}$) was reported for 4 pseudocubic LaInO$_{3}$ unit cells, and a decrease beyond it. We assign the differences to our predictions for non-periodic interfaces mainly to the high density of structural defects ({\it e.g}, dislocations) due to the large mismatch ($\sim 5\%$) between the channel and the substrate. The La doping, needed to compensate the acceptors induced by such dislocations, may cause an alleviation of the polar discontinuity at the interface and, hence, limit the 2DEG density. On the other hand, it makes the system metallic without a clear \textit{critical} thickness. With the recent advances in achieving high-quality BaSnO$_{3}$ and LaInO$_{3}$ single crystals~\cite{galazka+16jpcm,Zbigniew+20jcg,Aggoune+LIO} as well as interfaces~\cite{Martina+20prm,Martina+21}, our predictions open up a perspective for exploring interfacial charge densities in combinations of these materials in view of potential electronic applications. 

\textbf{Conclusions}

In summary, we have presented the potential of combining nonpolar BaSnO$_{3}$ and polar LaInO$_{3}$ for reaching a high interfacial carrier density. Our calculations show that, depending on the interface termination, both electron and hole gases can be formed that compensate the polar discontinuity. The gradual decrease of octahedra tilts from the orthorhombic LaInO$_{3}$ to the cubic BaSnO$_{3}$ side increases the out-of-plane lattice spacing at the interface and governs the unequal oxygen-cation displacements within the octahedra. The latter distortions induce a depolarization field, counteracting the formal polarization in the LaInO$_{3}$ block and hampering electronic reconstruction, {\it i.e.} the formation of a 2DEG at the interface up to a \textit{critical} LaInO$_{3}$ thickness of seven (five) pseudocubic unit cells for periodic (non-periodic) heterostructures. While the PBSEsol functional provides a good description of the interfacial charge-density distributions as well as the type of band offset, it fails to determine $t_{\mathrm{c}}$ reliably, as the knowledge of the quasiparticle gaps of the pristine materials is required. The polar distortions (polar discontinuity) decrease (increases) with LaInO$_{3}$ thickness, leading to a progressive charge transfer until reaching a 2DEG density of 0.5 e per surface unit cell. The electronic charge density is hosted by a highly dispersive Sn-\textit{s}-derived conduction band, suggesting a high carrier mobility. Overall, the 2DEG charge density can be tuned through the LaInO$_{3}$ thickness. Interestingly, also the type of interface ({\it i.e.}, periodic or non-periodic heterostructure) strongly impacts its density and spatial confinement. All these effects can be exploited in view of tailoring the characteristics of the 2DEG.

\section*{METHODS}
\textbf{Theory}

Ground-state properties are calculated using density-functional theory (DFT), within the generalized gradient approximation (GGA) in the PBEsol parameterization \cite{PBEsol+08prl} for exchange-correlation effects. All calculations are performed using FHI-aims~\cite{FHI-aims}, an all-electron full-potential package, employing numerical atom-centered orbitals. For all atomic species we use a \textit{tight} setting with a \textit{tier} 2 basis set for oxygen, \textit{tier1+fg} for barium, \textit{tier1+gpfd} for tin, \textit{tier 1+hfdg} for lanthanum, and \textit{tier 1+gpfhf} for indium. The convergence criteria are 10$^{-6}$ electrons for the density, 10$^{-6}$ eV for the total energy, 10$^{-4}$ eV \AA$^{-1}$~for the forces, and 10$^{-4}$ eV for the eigenvalues. Lattice constants and internal coordinates are optimized for all systems until the residual forces on each atom are less than 0.001 eV \AA$^{-1}$. The sampling of the Brillouin zone is performed with an 8 $\times$ 8 $\times$ 8 $\textbf{k}$-grid for bulk BaSnO$_{3}$ and a 6 $\times$ 6 $\times$ 4 $\textbf{k}$-grid for bulk LaInO$_{3}$. These parameters ensure converged total energies and lattice constants of 8 meV per atom and 0.001\AA,~respectively. 

For the heterostructures, a 6 $\times$ 6 $\times$ 1 $\textbf{k}$-grid is used. The in-plane lattice parameters are fixed to $\sqrt{2}a_{\mathrm{BSO}}$ ($a_{\mathrm{BSO}}$ being the bulk BaSnO$_{3}$ lattice spacing) (see supplementary figure 1). For the non-periodic systems, vacuum of about 150 \AA~is included and a dipole correction is applied in the [001] direction in order to prevent unphysical interactions between neighboring replica. In this case, we fix the first two BaSnO$_{3}$ unit cells to the bulk structure to simulate the bulk-like interior of the substrate, and relax the other internal coordinates. For computing the electronic properties, a 20 $\times$ 20 $\times$ 1 $\textbf{k}$-grid is adopted for all systems. This parameter ensures converged densities of states and electron/hole charge densities up to 0.01 $e$ per $a^{2}$. Atomic structures are visualized using the software package VESTA~\cite{momm-izum11jacr}.

\section*{Data availability}
Input and output files can be downloaded free of charge from the NOMAD Repository~\cite{drax-sche19jpm} at the following link: \url{https://dx.doi.org/10.17172/NOMAD/2021.03.10-1}.

\section*{Acknowledgments} 
This work was supported by the project BaStet (Leibniz Senatsausschuss Wettbewerb, No. K74/2017) and was performed in the framework of GraFOx, a Leibniz Science Campus, partially funded by the Leibniz Association. We acknowledge the North-German Supercomputing Alliance (HLRN) for providing HPC resources that have contributed to the research results reported in this paper (project bep00078). W. A. thanks Martin Albrecht, Martina Zupancic, and Toni Markurt (Leibniz-Institut f\"{u}r Kristallz\"{u}chtung, Berlin) Dmitrii Nabok, Le Fang, and Sebastian Tillack (Humboldt-Universit\"{a}t zu Berlin) as well as Kookrin Char (Seoul National University) for fruitful discussions. 

\section*{AUTHOR CONTRIBUTIONS}
W.A. performed the atomistic simulations and analyzed the data. All authors discussed the results and wrote the paper.
\section*{COMPETING INTERESTS}
The authors declare no competing interests.

\section*{Additional information}
Supplementary information is available for this paper at (/https:/...).

\newpage

\renewcommand{\figurename}{}
\renewcommand{\tablename}{}
\setcounter{figure}{0}   
\renewcommand{\thepage}{S\arabic{page}} 
\renewcommand{\thetable}{Supplementary table \arabic{table}}  
\renewcommand{\thefigure}{Supplementary figure \arabic{figure}} 
\newpage
\onecolumngrid

\newpage
{\centering
{\large 
\textbf{Supporting Information on}\\
\vskip 0.2cm
\textbf{"Tuning two-dimensional electron (hole) gases at LaInO$_{3}$/BaSnO$_{3}$ interfaces: Impact of polar distortions, termination, and thickness"}}\newline

Wahib Aggoune$^{1}$ and Claudia Draxl$^{1, 2}$\\
\textit{$^{1}$Institut f\"{u}r Physik and IRIS Adlershof, Humboldt-Universit\"{a}t zu Berlin, 12489 Berlin, Germany}\\
\textit{$^{2}$European Theoretical Spectroscopy Facility (ETSF)}\\

\vskip 0.5cm
\date{today} 
}

 \twocolumngrid
   \normalsize

 
\section*{Supplementary Methods} 
  
\textbf{Local density of states and charge densities}

The local density of states per unit cell (LDOS) in the out-of-plane direction is obtained by summing over the atom-projected density of states of the atoms within the respective unit cell. The density of the two-dimensional electron gas (2DEG) is evaluated by integrating the LDOS of the occupied states between the conduction-band minimum and the Fermi level (E$_F$) and summing over all planes \textit{i.e.} AO (A = La, Ba) and BO$_{2}$ (B = Sn, In). Likewise, the hole-gas density (2DHG) is obtained from the corresponding integral over the unoccupied states between the E$_F$ and valence-band maximum. For the non-stoichiometric nn- and pp-type periodic heterostructures, we integrate over the states comprised between the mid gap and the Fermi level. 
  
\section*{Supplementary Discussion} 

\textbf{Structural properties}

\begin{figure}
 \begin{center}
\includegraphics[width=.45\textwidth]{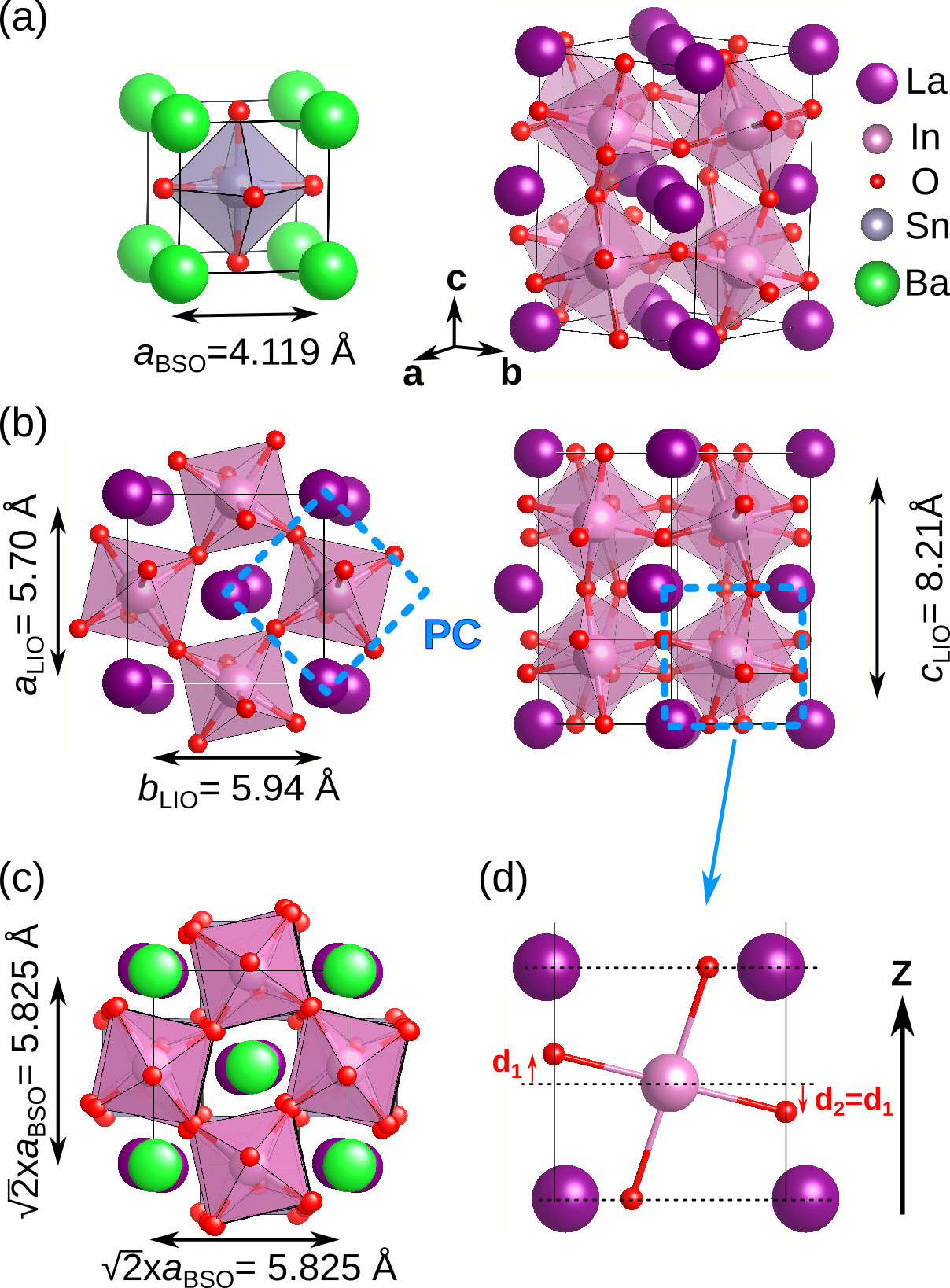}%
\caption{\textbf{Structural properties of pristine materials.} (a) Primitive cell of cubic BaSnO$_{3}$ (left) and orthorhombic LaInO$_{3}$ (right). (b) Top and side view of the LaInO$_{3}$ unit cell. The corresponding pseudocubic unit cell (PC) is highlighted by dashed light-blue lines. (c) Top view of the heterostructure where the in-plane lattice dimensions are fixed to the bulk BaSnO$_{3}$ geometry. (d) Equal out-of-plane displacements (with opposite sign) of the inequivalent oxygen atoms with respect to In by tilting the octahedra.}
\label{fig:bulk}
 \end{center}
\end{figure}

The primitive unit cell in BaSnO$_{3}$ contains 5 atoms, with highly symmetric (non-tilted) SnO$_{6}$ octahedra [see \ref{fig:bulk}(a)]. The pseudocubic LaInO$_{3}$ unit cell is defined as the structure exhibiting the same volume per LaInO$_{3}$ formula unit as the orthorhombic cell [see \ref{fig:bulk}(b)]. The calculated averaged pseudocubic lattice parameter is 4.116~\AA. In \ref{fig:bulk}(c) we show the top view of an interface where the in-plane lattice parameters are fixed to $\sqrt{2}a_{\mathrm{BSO}}$ ($a_{\mathrm{BSO}}$ being the bulk BaSnO$_{3}$ lattice spacing). Details on the structural parameters of the considered heterostructures are summarized in Table~\ref{tab:str}.

\begin{table*}[!htbp]
\centering
\caption{Structural parameters of the heterostructure geometries shown in Fig. 1 (stoichiometric periodic LIO$_{12}$/BSO$_{10}$), Fig. 2 (non-stoichiometric periodic nn-type LIO$_{10}$/BSO$_{10}$), and Fig. 3 (stoichiometric non-periodic LIO$_{8}$/BSO$_{11}$), shown in the main text, as well as in Supplementary figure 5 (non-stoichiometric periodic pp-type LIO$_{10}$/BSO$_{10}$). The non-stoichiometric nn- and pp-type systems are symetric and have overall nonpolar character.}

\vspace{0.2cm}
 \begin{tabular}{|c|c|c|c|c|c|}  
 \hline
Superlattices &  Formula & Spacegroup &a=b=$\sqrt{2}~a_{\mathrm{BSO}}$ [\AA]&c [\AA]\\
\hline
Stoichiometric LIO$_{12}$/BSO$_{10}$ &Ba$_{20}$Sn$_{20}$La$_{24}$In$_{24}$O$_{132}$ & (7) P c &5.825&90.664\\
nn-type LIO$_{10}$/BSO$_{10}$ & Ba$_{18}$Sn$_{20}$La$_{22}$In$_{20}$O$_{120}$  & (26) P m c 2$_{1}$ &5.825&82.498\\
pp-type LIO$_{10}$/BSO$_{10}$ & Ba$_{22}$Sn$_{20}$La$_{18}$In$_{20}$O$_{120}$    &(26) P m c 2$_{1}$ &5.825&82.449  \\
\hline
\hline
Interface (with 150 \AA~vacuum) &  Formula & Spacegroup&a=b=$\sqrt{2}~a_{\mathrm{BSO}}$ [\AA]&c [\AA]\\
\hline
Stoichiometric LIO$_{8}$/BSO$_{11}$ &Ba$_{22}$Sn$_{22}$La$_{16}$In$_{16}$O$_{114}$  & (1) P 1 &5.825&226.283\\
\hline
\end{tabular}
\label{tab:str}
\end{table*}

\begin{figure*}
 \begin{center}
\includegraphics[width=.99\textwidth]{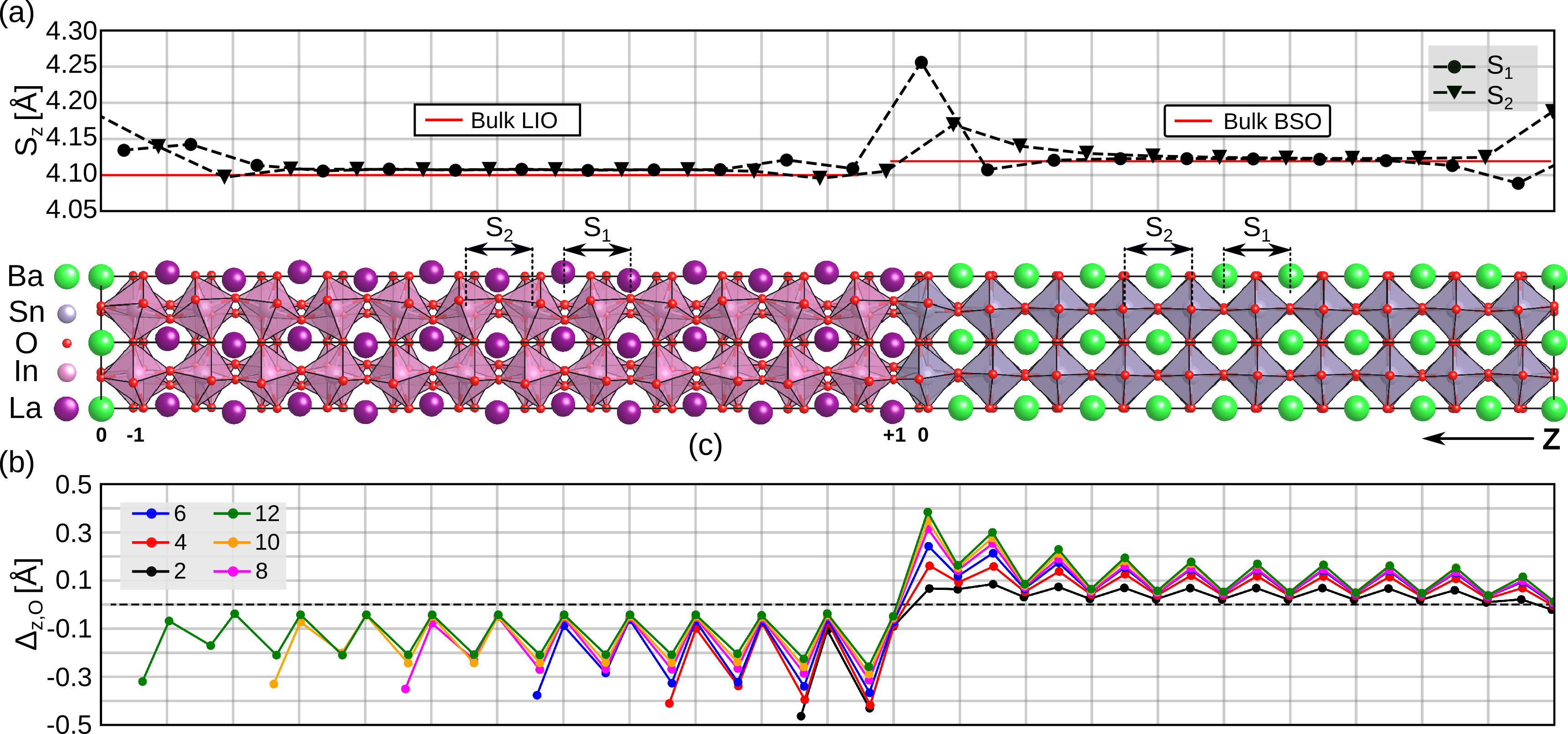}%
\caption{\textbf{Properties of the stoichiometric periodic heterostructure LIO$_{12}$/BSO$_{10}$.} (a) Out-of-plane spacing S$_{1}$ (S$_{2}$) between adjacent Ba (Sn) atoms within BaSnO$_{3}$ and between adjacent La (In) atoms within the LaInO$_{3}$ block, also indicated by arrows in the structure model below. The red lines indicate the spacing in the bulk counterparts (b) Oxygen displacements $\Delta _{z,\mathrm{O}}$ within each unit cell for different LaInO$_{3}$ thicknesses (in units of the pseudocubic unit cell). The lines are guides to the eye. For details see \ref{fig:cell-def}.}
\label{fig:O-disp-LaIn}
 \end{center}
\end{figure*}

As shown in \ref{fig:O-disp-LaIn}(a) for the stoichiometric periodic heterostructure LIO$_{12}$/BSO$_{10}$, the out-of-plane lattice spacing increases locally due to the gradual change of the octahedra tilts, amounting to about +3\% with respect to that of bulk BaSnO$_{3}$ at the interface. This finding is in very good agreement with experimental observations~\cite{Markurt+19sr}, reporting a local increase of about 2\%. We also clearly see that 10 unit cells of BaSnO$_{3}$ are enough to minimize the interaction with its replica, since the out-of-plane lattice spacing in the middle of BaSnO$_{3}$ block is similar to its bulk counterpart [\ref{fig:O-disp-LaIn}(a)].\\

\begin{figure*}
 \begin{center}
\includegraphics[width=.6\textwidth]{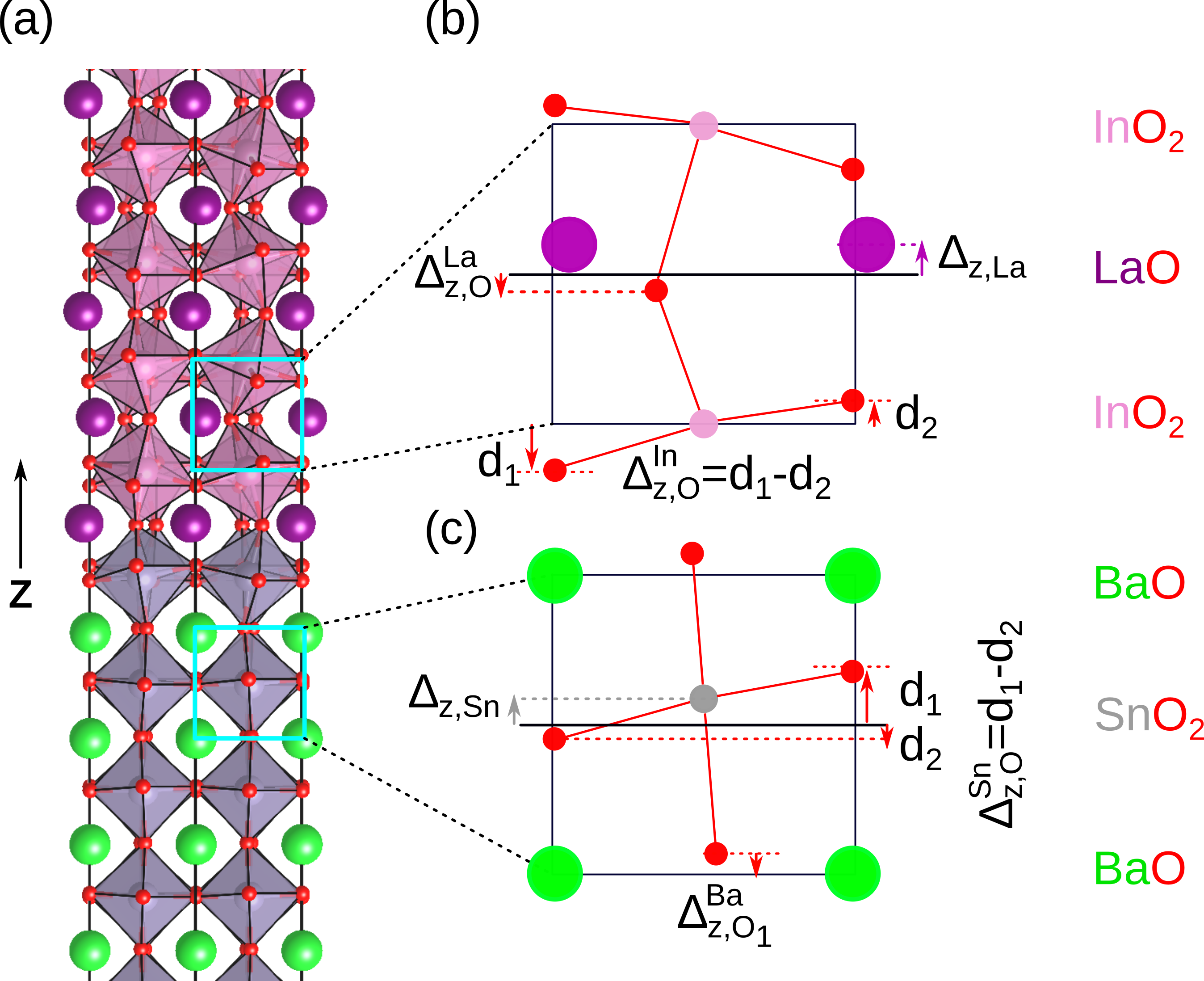}%
\caption{\textbf{Structural distortions within the stoichiometric periodic heterostructure.} (a) Structural model of LIO$_{2}$/BSO$_{10}$ system. (b) Schematic showing the out-of-plane displacements $\Delta_{z}$ of the different ions within LaInO$_{3}$. The unit cell covers the space between adjacent InO$_{2}$ planes, taking In as the reference. $\Delta _{z,\mathrm{La}}$ and $\Delta _{z,\mathrm{O}}^{\mathrm{La}}$ represent the displacements from the bulk $z$ coordinate. $\Delta _{z,\mathrm{O}}^{\mathrm{In}}$ represents the difference between the displacements $d_{1}$ and $d_{2}$ of the inequivalent O atoms with respect to the cation In. (c) Same as (b) for the ions within BaSnO$_{3}$. In this case, the unit cell is defined as the space between adjacent BaO planes, taking Ba as the reference. $\Delta _{z,\mathrm{O}}^{\mathrm{Ba}}$ represents the distance from its bulk $z$ coordinate. $\Delta _{z,\mathrm{O}}^{\mathrm{Sn}}$ represents the difference between the displacements $d_{1}$ and $d_{2}$ of the inequivalent O atoms along z. $\Delta _{z,\mathrm{Sn}}$ is the displacement of the Sn atom from its bulk $z$ coordinate.}

\label{fig:cell-def}
 \end{center}
\end{figure*}

\textbf{Polarization}

A qualitative trend of the polarization induced by the structural distortions are estimated using the relative atomic displacements in the supercell together with the Born effective charges ($Z^{*}$) calculated for bulk BaSnO$_{3}$ and LaInO$_{3}$, along the $z$ direction. Such approach has been largely used in literature for the description of polarization effects in oxides interfaces~\cite{Pickett+09prl}. The fact that it cannot be fully quantitative, is attributed to local dipoles induced in the heterostructures, which do not exist in the bulk, as well as to the possible appearance of a metallic character across the interface. 
As a consequence, polarization and 2DEG densities are not the same in the two models. However, the periodic heterostructures a nevertheless a valuable approach for analyzing trends, like how the polarization behaves with the thickness of the polar LaInO$_{3}$ block. 

$Z^{*}$ are computed within the Berry phase approach~\cite{berryPhase+93prbr} using \texttt{exciting}~\cite{gula+14jpcm}, an all-electron full-potential code, implementing the family of (L)APW+LO (linearized augmented planewave plus local orbital) methods. These calculations are performed for the relaxed bulk structures of BaSnO$_{3}$ and LaInO$_{3}$ (\ref{fig:bulk}) as obtained by the FHI-aims code, and using the same $\textbf{k}$-grid. For the atomic species A (Ba, La), B (Sn, In), and O, muffin-tin radii (R$_{MT}$) of 2.2, 2.0, and 1.6 bohr are used, respectively. A basis-set cutoff R$_{MT}$G$_{max}$=8 is adopted, where R$_{MT}$ here refers to the radius of the smallest sphere (1.6 bohr), {\it i.e.}, G$_{max}$=5. 

The results for Ba, Sn, O within the BaO layer, and O within the SnO$_{2}$ layer are $Z^{*}$=2.77, 4.44, -3.47, and -1.87, respectively. For LaInO$_{3}$, we obtain 3.83, 3.38, -2.53, and -2.34, for La, In, O within the LaO layer, and O within the InO$_{2}$ layer, respectively. As depicted in \ref{fig:bulk} (d), the InO$_{2}$ octahedra in bulk LaInO$_{3}$ are tilted, and thus the two inequivalent oxygen atoms are shifted up and down ($z$ direction), respectively, with respect to the In atom by the same amount. Likewise, the inequivalent La atoms within different planes are displaced within the same plane but in opposite direction, away from their centrosymmetric positions. Consequently, the induced polarizations arising from the respective inequivalent atoms cancel each other in bulk LaInO$_{3}$, resulting in zero total polarization [see \ref{fig:bulk}(d)]. This means that the tilting of octahedra does not necessarily induce a polarization. Such octahedra tilts are thus termed nonpolar distortions~\cite{Gazquez+17prl}. They can, however, induce a polarization in the out-of-plane direction when the oxygen displacements are unequal, {\it i.e.}, $\Delta _{z,\mathrm{O}}=d_{1}-d_{2}\neq 0$. This situation is indeed observed in our heterostructures [see the example in~\ref{fig:cell-def} and~\ref{fig:nn-vs-pp}(c)].

Using a simple linear approximation~\cite{Resta+93prl}, the local polarization per unit cell is written as
\begin{equation}
P_{z}=\frac{1}{\Omega}\sum_{i}Z^{*}_{z,i}\Delta _{z,i}
\label{eq:P}
\end{equation} 
where, $\Omega$ is the volume of the unit cell and $ Z^{*}_{z,i}$ is the Born effective charge of ion $i$ along the \textit{z} direction. For the choice of unit cells and definitions of the displacements $\Delta _{z}$ of the different ions within LaInO$_{3}$ and BaSnO$_{3}$ see \ref{fig:cell-def}.

In \ref{fig:O-disp-LaIn}(b), we display the variation of the oxygen distortions, $\Delta _{z,\mathrm{O}}$, along the considered layers for the stoichiometric periodic heterostructures with composition LIO$_{m}$/BSO$_{10}$ ($m$ = 2, 4, 6, 8, 10, 12). As we can see, the O atoms are shifted up (\textit{z} direction) within the BaSnO$_{3}$ side and down within the LaInO$_{3}$ side. As $ Z^{*}_{z,\mathrm{O}}$ is negative, the induced polarization in the BaSnO$_{3}$ (LaInO$_{3}$) side is opposite to (the same as) the out-of-plane direction. In the LaInO$_{3}$ side, it counteracts the formal polarization, while in the BaSnO$_{3}$ side it leads to a minimization of the polar discontinuity at the interface. The distortions $\Delta _{z,\mathrm{O}}$ in the BaSnO$_{3}$ side increase with the LaInO$_{3}$ thickness, while they decrease in the LaInO$_{3}$ side. Note that $\Delta _{z,\mathrm{O}}$ has the main contribution to the polar distortions shown in Figs.~1 (c) and (g) of the main text.

\begin{figure*}
 \begin{center}
\includegraphics[width=.78\textwidth]{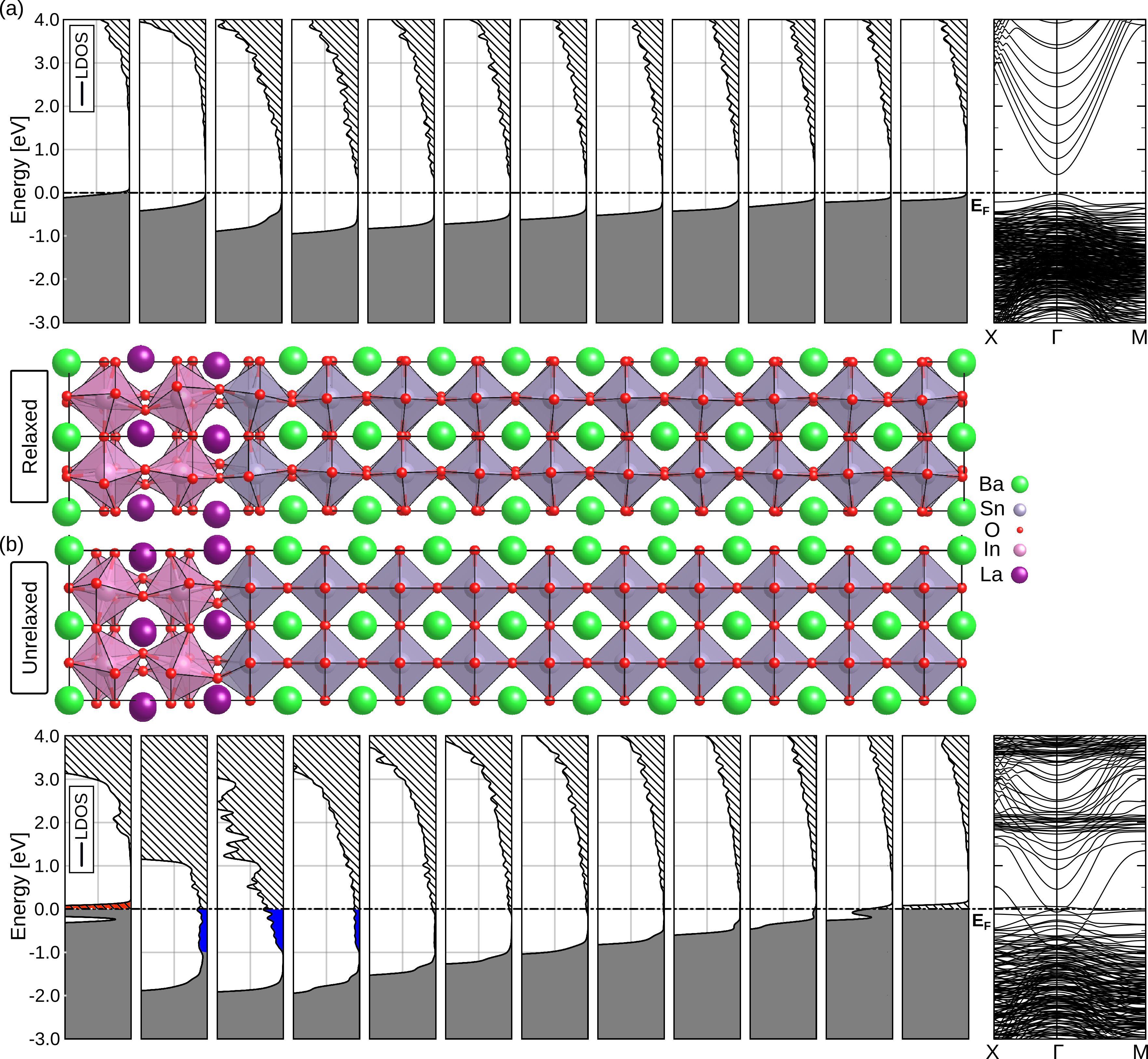}%
\caption{\textbf{Comparison between} the \textit{relaxed} (a) and the \textit{unrelaxed} (b) stoichiometric periodic heterostructure LIO$_{2}$/BSO$_{10}$. The LDOS (left) and the band structure (right) show the semiconducting character of the former and the metallic character of the latter system.}
\label{fig:relaxed-vs-unrelaxed}
 \end{center}
\end{figure*}

The highest value of polarization induced by these distortions is found for the periodic heterostructure LIO$_{2}$/BSO$_{10}$ [see \ref{fig:relaxed-vs-unrelaxed} and Fig.~1(g) in the main text]. Counteracting the formal polarization $P_{0}$, it hampers the electronic charge transfer up to a thickness of 4 pseudocubic LaInO$_{3}$ unit cells [Fig.~1(f) in the main text]. Ignoring these distortions by considering \textit{unrelaxed} periodic heterostructure, we find that it has a metallic character with a complete electronic reconstruction [see \ref{fig:relaxed-vs-unrelaxed}(b)]. Consequently, a 2DEG of about 0.5 $e$ per $a^{2}$ is formed as predicted by the polar-catastrophe model that is based on a rigid, unrealxed lattice. This result emphasizes the critical role of polar distortions in compensating the polar discontinuity. Reaching the \textit{critical} thickness of 4 pseudocubic LaInO$_{3}$ unit cells, a 2DEG forms inside BaSnO$_{3}$. Increasing the LaInO$_{3}$ thickness, the charge density increases and the polar distortions decrease. The latter behavior implies that the octahedral tilt within the LaInO$_{3}$ converges to that of the bulk counterpart upon increasing the thickness. The same behavior is found for the non-periodic systems, but with less pronounced structural distortions in the BaSnO$_{3}$ substrate [see also Fig.~3 (c), (f) and (g) in the main text].\\
 
\textbf{Comparison between n- and p-type interfaces} 
\begin{figure*}[t]
 \begin{center}
\includegraphics[width=.98\textwidth]{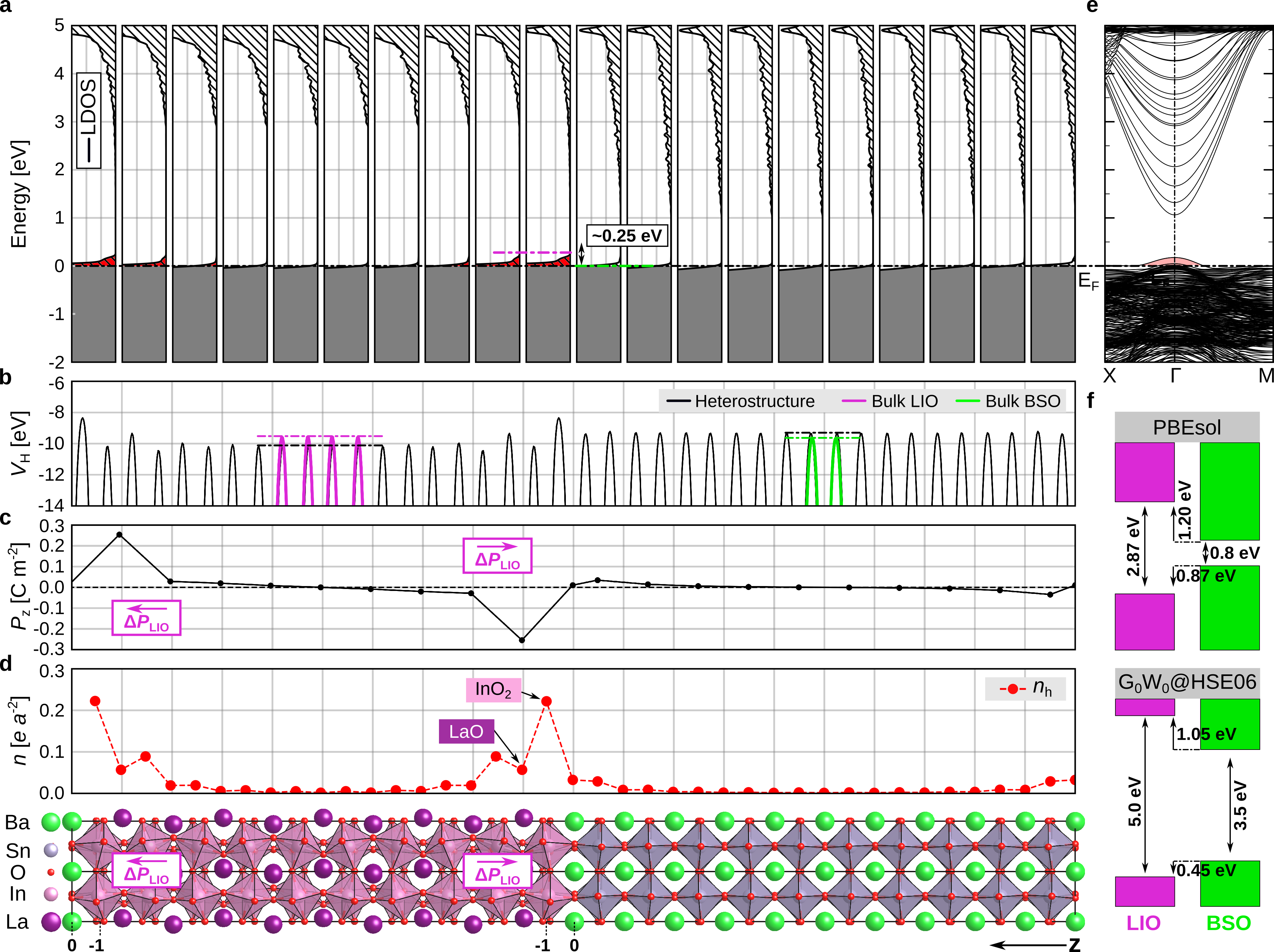}%
\caption{\textbf{Non-stoichiometric pp-type periodic heterostructure:} the system is formed by LIO$_{10}$/BSO$_{10}$ in out-of-plane direction \textit{z} (bottom), that is shared between panels (a), (b), (c), and (d) as their horizontal axis. The LaInO$_{3}$ block is terminated by a (InO$_{2}$)$^{-1}$ plane on both sides. (a) Local density of states per unit cell (LDOS). The Fermi level is set to zero, and the unoccupied states of the (original) valence bands are highlighted by the shaded red area (also in panel (e)). The dashed green and pink lines indicate the alignment of the valence-band edges of BaSnO$_{3}$ and LaInO$_{3}$ at the interface. (b) In-plane averaged electrostatic potential along the \textit{z} direction. (c) Total polarization per unit cell. (d) Distribution of the hole charge density obtained by integrating the DOS over the unoccupied valence states indicated by the red color in panel (a). (e) Electronic band structure along X-$\Gamma$-M. (f) Band alignment as obtained by using the band gaps of the bulk systems (PBEsol [top] and G$_{0}$W$_{0}$@HSE06 [bottom]), considering the energy difference between the potential of the bulk and the periodic heterostructure as shown in panel (b). The alignment at the interface obtained by PBEsol can be seen in panel (a).}
\label{fig:InIn-HS}
 \end{center}
\end{figure*}

In \ref{fig:InIn-HS}, we present the electronic properties of a non-stoichiometric periodic heterostructure, where the LaInO$_{3}$ slab terminates with an (InO$_{2}$)$^{-1}$ plane on both sides (termed pp-type). The electronic band structure shows that this system has metallic character, where the the (former) valence band in LaInO$_{3}$ is partially depleted by one $e$ per $a^{2}$ [see \ref{fig:InIn-HS}(e)]. The effective hole mass is about 1.21 m$_{e}$, significantly higher than that of bulk LaInO$_{3}$ (0.48 m$_{e}$)~\cite{Aggoune+LIO}. The octahedral tilt decreases gradually from the LaInO$_{3}$ to the BaSnO$_{3}$ side. However, in this case the BaO termination favors nonpolar distortions within the BaSnO$_{3}$ side ({\it i.e.}, equal displacements of the inequivalent O atoms). Consequently, polar distortions appear mainly within the LaInO$_{3}$ side, inducing polarizations oriented toward the interface. This local dipole causes an upward shift of the valence-band edge inside LaInO$_{3}$ near the interface as also seen in the averaged in-plane electrostatic potential [see \ref{fig:InIn-HS}(b)]. The valence-band edge at the BaSnO$_{3}$ side is almost the same in all unit cells as polar distortions are negligible [see~\ref{fig:InIn-HS}(a) and (c)]. Integrating the LDOS over the partially unoccupied valence states, we find that the 2DHG stays in the LaInO$_{3}$ side of the interface and is confined within one pseudocubic unit cell, amounting to 0.5 $e$ per $a^{2}$ at each side [see \ref{fig:InIn-HS}(d)]. It is mainly hosted by O-\textit{p} states within the InO$_{2}$ layers.

In \ref{fig:nn-vs-pp}, we compare the optimized geometries of nn- and pp-type periodic heterostructures. Before analyzing the latter, we note that in a pristine symmetric LaInO$_{3}$ slab with InO$_{2}^{-1}$ termination, the hole charge accumulates on its surfaces, accompanied by structural distortions that tend to screen the discontinuity of the polarization. Combined with BaSnO$_{3}$ (pp-type periodic heterostructure), the hole charge stays at the LaInO$_{3}$ side. The structural distortions also appear only in the LaInO$_{3}$ side [see \ref{fig:nn-vs-pp}(b)]. Thus, the interfacial polar discontinuity is less compensated than in the n-type interface, making the latter more favorable. This explains why, at the interface, (n-type) SnO$_{2}$ termination is predominantly found experimentally~\cite{Martina+21}. This is also confirmed by computing the formation energies, that differ by 0.05 eV per atom (-2.40 eV per atom compared to -2.45 eV per atom). \\

\begin{figure*}
 \begin{center}
\includegraphics[width=.98\textwidth]{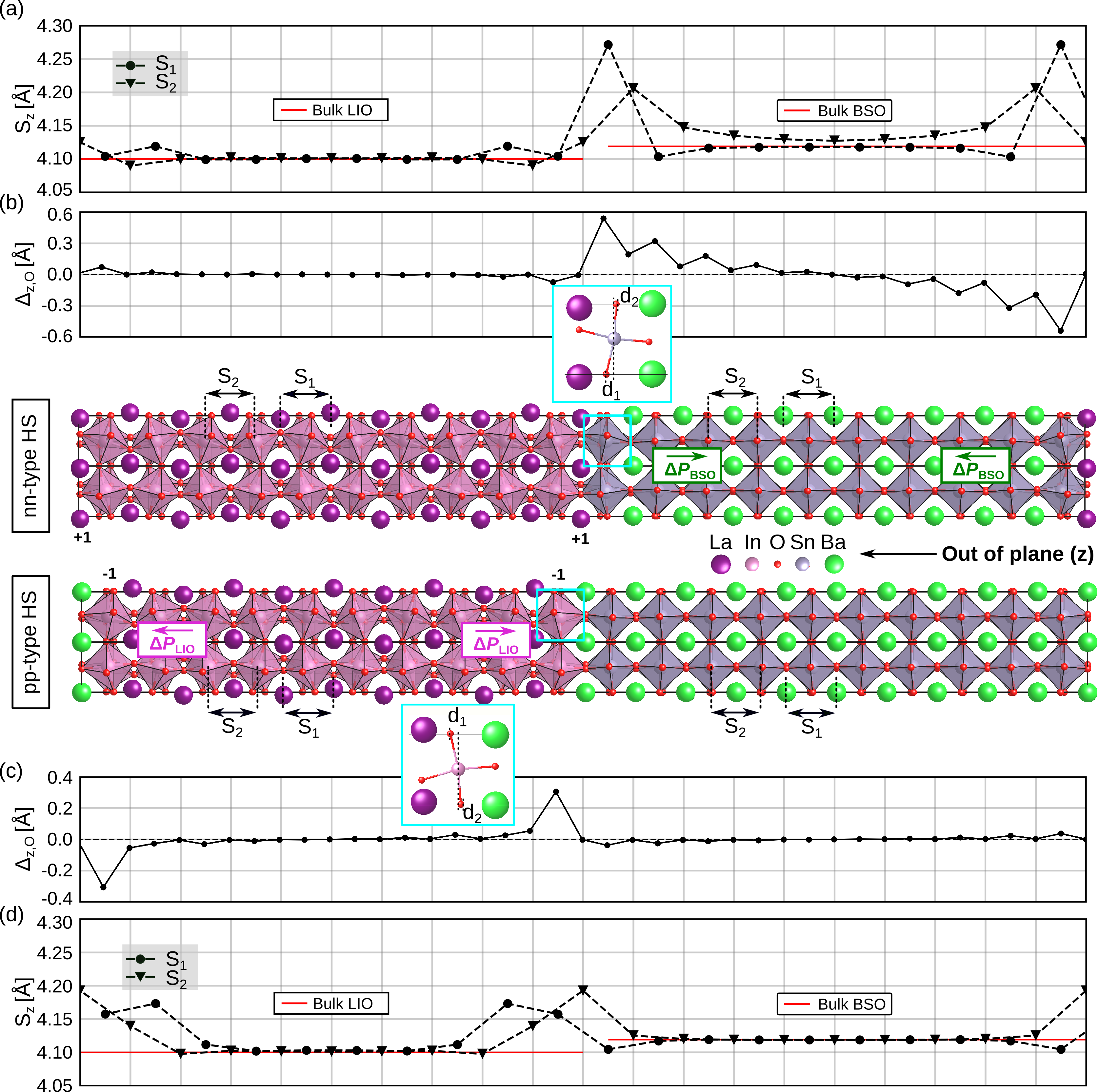}%
\caption{\textbf{Properties of the non-stoichiometric periodic heterostructure.} Out-of-plane spacing S$_{1}$ (S$_{2}$) between adjacent Ba (Sn) atoms within BaSnO$_{3}$ and between adjacent La (In) atoms within the LaInO$_{3}$ block, also indicated by arrows in the structure model for the (a) nn-type and (d) pp-type periodic heterostructure. The red lines indicate the spacing in the bulk counterparts. Displacement of oxygen atoms $\Delta _{z,\mathrm{O}}$ in the (b) nn-type and (c) pp-type models. For details see \ref{fig:cell-def}.}
\label{fig:nn-vs-pp}
 \end{center}
\end{figure*}
\textit{Band alignment in non-stoichiometric periodic heterostructures}

To compute the band offset using the quasiparticle band gaps of the bulk materials BaSnO$_{3}$ and LaInO$_{3}$~\cite{Aggoune+BSO,Aggoune+LIO}, we use an approach based on the electrostatic potential that accounts for all electrostatic effects at the interface, {\it i.e.}, charge rearrangements upon interface formation and interface dipoles~\cite{dima+19cm}. We first estimate the energy shift between the in-plane averaged electrostatic potential of the pristine components (orthorhombic LaInO$_{3}$ and cubic BaSnO$_{3}$) and that of the periodic heterostructure [see Fig.~2(b) in the main text and \ref{fig:InIn-HS}(b)]. This energy difference is used to shift the valence- and conduction-band edges calculated for the bulk materials in order to determine the corresponding band alignment in the heterostructure. We note that this approach provides the alignment only away from the interface as the potential-energy shift can be only estimated there [see Fig.~2(b) in the main text and \ref{fig:InIn-HS}(b)]. We find that the conduction-band offset between BaSnO$_{3}$ and LaInO$_{3}$ is almost the same using the PBEsol gaps or the respective quasiparticle gaps~\cite{Aggoune+BSO,Aggoune+LIO} [see Fig.~2(f) in the main text]. Likewise, for the periodic pp-type heterostructure, a similar valence-band offset is found using PBEsol or the quasiparticle gaps [see \ref{fig:InIn-HS}(f)]. We note that the band alignment exactly at interface can be evaluated from the LDOS only [see Figs.~1(a), 2(a), 3(a) in the main text and \ref{fig:InIn-HS}(a)]. Overall, these results show that the PBEsol functional is good enough to capture band offset and charge distribution at the interface.\\

\textit{HSE06 calculation for validating the scissor-shift approach}

\begin{figure}
 \begin{center}
\includegraphics[width=.46\textwidth]{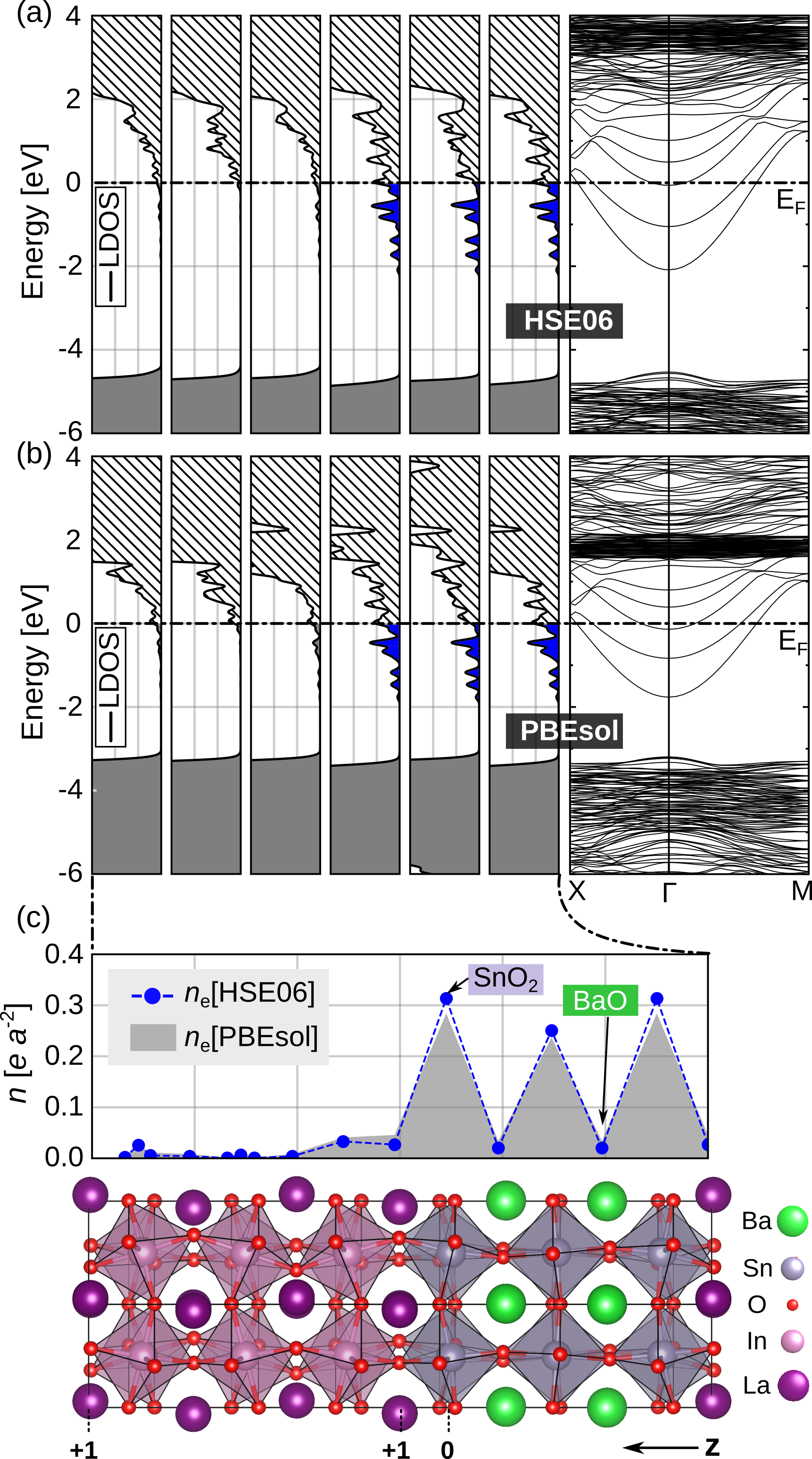}%
\caption{\textbf{Non-stoichiometric periodic nn-type heterostructure:} the system is formed by LIO$_{3}$/BSO$_{3}$ in the out-of-plane direction \textit{z} (bottom), that is shared between panels (a), (b), and (c) as their horizontal axis. Local density of states per unit cell (LDOS) obtained using (a) HSE06 and (b) PBEsol. The Fermi level is set to zero, and the electron population of the (original) conduction bands is shown as shaded blue area. The corresponding band structures along X-$\Gamma$-M are shown in the right panels. (c) Distribution of the electron charge density obtained by integration over the occupied conduction states indicated by the blue area in panel (a). The electronic distribution obtained by PBEsol (from panel b) is shown for comparison.}

\label{fig:hse-vs-pbesol}
 \end{center}
\end{figure}

To justify the scissor-shift approach considered above, we compute the electronic properties of a feasible heterostructure using HSE06 for comparison with PBEsol. For more information on its performance for oxide perovskites, we refer to calculations for pristine BaSnO$_{3}$ and LaInO$_{3}$ materials~\cite{Aggoune+BSO,Aggoune+LIO}.  In \ref{fig:hse-vs-pbesol}, we depict the results for a non-stoichiometric nn-type periodic heterostructure, formed by three BaSnO$_{3}$ unit cells and three pseudocubic LaInO$_{3}$ unit cells. Comparing panels (a) and (b), we clearly see that the LDOS obtained by HSE06 and PBEsol, respectively, have similar band alignment and shape within each unit cell. Form the right panels, we also see that the band edges are similar. Differences only appear in higher-lying energy bands (between 2 and 4 eV), attributed to the fact that these bands are formed by localized Ba-\textit{d} and -\textit{f} states as well as La-\textit{f} orbitals which are better described by HSE06.  Overall, the conduction-band edge is well captured by PBEsol. This can be attributed to the fact that the valence- (conduction-)band edge of both materials is made of delocalized \textit{p}-states (\textit{s}-states). This finding validates the \textit{critical} thickness of 7 pseudocubic LaInO$_{3}$ unit cells estimated in the main text for the stoichiometric periodic heterostructure by relying on the offset given by PBEsol [LDOS, Fig.~1(a) in the main text] that is corrected by a scissor shift according to the quasiparticle gaps of the pristine materials~\cite{Aggoune+BSO,Aggoune+LIO}. Applying the same procedure, for the non-periodic interface, a \textit{critical} thickness of five pseudocubic LaInO$_{3}$ unit cells is found. \\

\textbf{Comparison between periodic and non-periodic systems}

In the main text, we show that, for a given LaInO$_{3}$ thickness, the 2DEG density is higher in the non-periodic heterostructures compared to periodic ones [see Fig.~3(f)]. As shown in \ref{fig:HS-vs-IF} for a thickness of eight pseudocubic LaInO$_{3}$ unit cells, the enhanced 2DEG charge density in the non-periodic system is due to the pronounced polar discontinuity at the interface. The latter is attributed to smaller polar distortions within the BaSnO$_{3}$ side. This, in turn, allows for an extension of the 2DEG up to five unit cells into the BaSnO$_{3}$ substrate. In the periodic heterostructures, the significant polar distortions in the BaSnO$_{3}$ side reduce the interfacial polar discontinuity and thus the electronic charge density. However, such distortions lead to a confinement of the 2DEG within three BaSnO$_{3}$ unit cells compared to five in the non-periodic structure.

\begin{figure*}
 \begin{center}
\includegraphics[width=.98\textwidth]{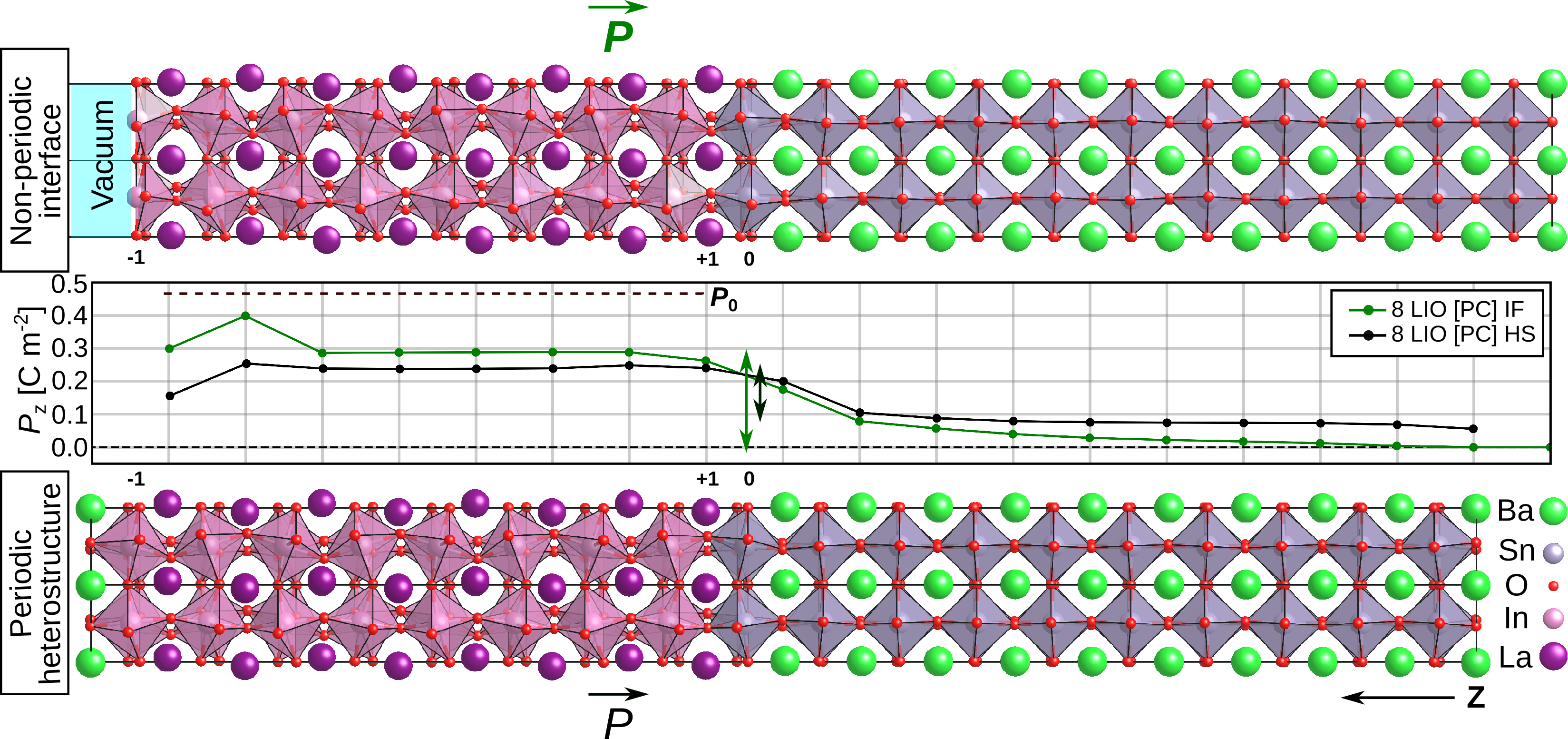}%
\caption{\textbf{Comparison between the polar discontinuity} in the stoichiometric, periodic (termed HS) and non-periodic (termed IF) heterostructures both comprising eight pseudocubic LaInO$_{3}$ unit cells. Total polarization in the out-of-plane direction for the non-periodic (green) and periodic (black) heterostructure. The dashed black line indicates the formal polarization $P_{0}$. The vertical arrows highlight the interfacial polar discontinuity.}
\label{fig:HS-vs-IF}
 \end{center}
\end{figure*}

\begin{figure*}
 \begin{center}
\includegraphics[width=.98\textwidth]{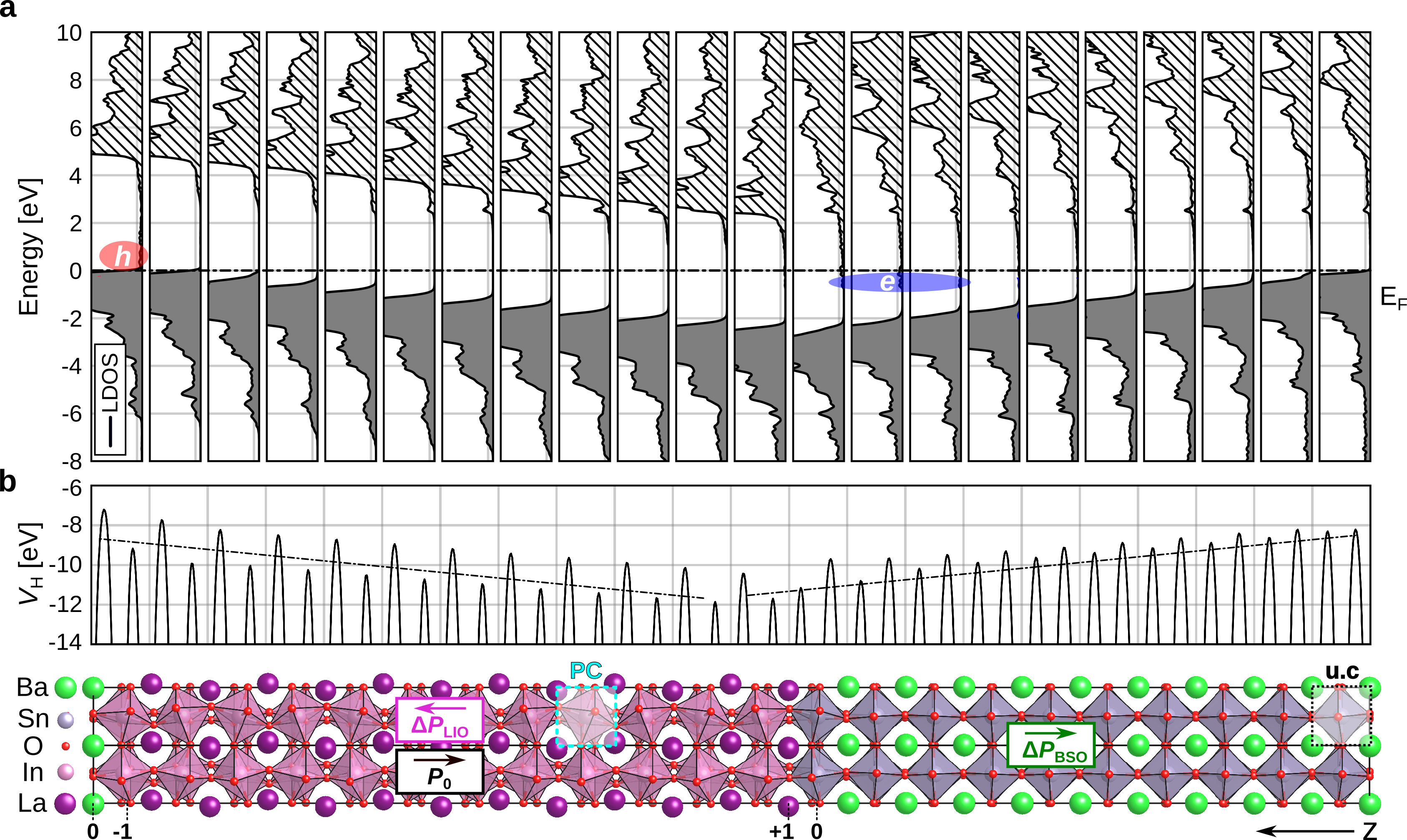}%
\caption{(a) Local density of states per unit cell (LDOS) of the stoichiometric periodic LIO$_{12}$/BSO$_{10}$ heterostructure showing higher energy range compared with Fig. 1 in the main text, in order to provide a complete trend of the valence and conduction band. (b) In-plane averaged electrostatic potential along the \textit{z} direction. The dashed lines are guides to the eye, highlighting the trend of the potential.
}
\label{fig:HS-vs-IF}
 \end{center}
\end{figure*}

\section*{Supplementary References}


\section*{References}

\begin{thebibliography}{10}
\expandafter\ifx\csname url\endcsname\relax
  \def\url#1{\texttt{#1}}\fi
\expandafter\ifx\csname urlprefix\endcsname\relax\def\urlprefix{URL }\fi
\providecommand{\bibinfo}[2]{#2}
\providecommand{\eprint}[2][]{\url{#2}}

\bibitem{Ohtomo+04n}
\bibinfo{author}{Ohtomo, A.} \& \bibinfo{author}{Hwang, H.~Y.}
\newblock \bibinfo{title}{A high-mobility electron gas at the
  $\mathrm{LaAlO_{3}/SrTiO_{3}}$ heterointerface}.
\newblock \emph{\bibinfo{journal}{Nature}} \textbf{\bibinfo{volume}{427}},
  \bibinfo{pages}{423} (\bibinfo{year}{2004}).

\bibitem{Mannhart+10sc}
\bibinfo{author}{Mannhart, J.} \& \bibinfo{author}{Schlom, D.~G.}
\newblock \bibinfo{title}{Oxide interfaces{\textemdash}an opportunity for
  electronics}.
\newblock \emph{\bibinfo{journal}{Science}} \textbf{\bibinfo{volume}{327}},
  \bibinfo{pages}{1607--1611} (\bibinfo{year}{2010}).

\bibitem{Bert+11np}
\bibinfo{author}{Bert, J.~A.} \emph{et~al.}
\newblock \bibinfo{title}{Direct imaging of the coexistence of ferromagnetism
  and superconductivity at the $\mathrm{LaAlO_{3}/SrTiO_{3}}$ interface}.
\newblock \emph{\bibinfo{journal}{Nat. Phys.}} \textbf{\bibinfo{volume}{7}},
  \bibinfo{pages}{767} (\bibinfo{year}{2011}).

\bibitem{Nakagawa+06nm}
\bibinfo{author}{Nakagawa, N.}, \bibinfo{author}{Hwang, H.~Y.} \&
  \bibinfo{author}{Muller, D.~A.}
\newblock \bibinfo{title}{Why some interfaces cannot be sharp}.
\newblock \emph{\bibinfo{journal}{Nat. Mater.}} \textbf{\bibinfo{volume}{5}},
  \bibinfo{pages}{204} (\bibinfo{year}{2006}).

\bibitem{Huijben+09am}
\bibinfo{author}{Huijben, M.} \emph{et~al.}
\newblock \bibinfo{title}{Structure–property relation of
  $\mathrm{SrTiO_{3}/LaAlO_{3}}$ interfaces}.
\newblock \emph{\bibinfo{journal}{Adv. Mater.}} \textbf{\bibinfo{volume}{21}},
  \bibinfo{pages}{1665--1677} (\bibinfo{year}{2009}).

\bibitem{Chambers2+10ssr}
\bibinfo{author}{Chambers, S.} \emph{et~al.}
\newblock \bibinfo{title}{Instability, intermixing and electronic structure at
  the epitaxial $\mathrm{LaAlO_{3}/SrTiO_{3}}$(001) heterojunction}.
\newblock \emph{\bibinfo{journal}{Surf. Sci. Rep.}}
  \textbf{\bibinfo{volume}{65}}, \bibinfo{pages}{317 -- 352}
  (\bibinfo{year}{2010}).

\bibitem{Thiel+06sc}
\bibinfo{author}{Thiel, S.}, \bibinfo{author}{Hammerl, G.},
  \bibinfo{author}{Schmehl, A.}, \bibinfo{author}{Schneider, C.~W.} \&
  \bibinfo{author}{Mannhart, J.}
\newblock \bibinfo{title}{Tunable quasi-two-dimensional electron gases in oxide
  heterostructures}.
\newblock \emph{\bibinfo{journal}{Science}} \textbf{\bibinfo{volume}{313}},
  \bibinfo{pages}{1942--1945} (\bibinfo{year}{2006}).

\bibitem{Thiel+09prl}
\bibinfo{author}{Thiel, S.} \emph{et~al.}
\newblock \bibinfo{title}{Electron scattering at dislocations in
  $\mathrm{LaAlO_{3}/SrTiO_{3}}$ interfaces}.
\newblock \emph{\bibinfo{journal}{Phys. Rev. Lett.}}
  \textbf{\bibinfo{volume}{102}}, \bibinfo{pages}{046809}
  (\bibinfo{year}{2009}).

\bibitem{Xie+13am}
\bibinfo{author}{Xie, Y.}, \bibinfo{author}{Bell, C.}, \bibinfo{author}{Hikita,
  Y.}, \bibinfo{author}{Harashima, S.} \& \bibinfo{author}{Hwang, H.~Y.}
\newblock \bibinfo{title}{Enhancing electron mobility at the
  $\mathrm{LaAlO_{3}/SrTiO_{3}}$ interface by surface control}.
\newblock \emph{\bibinfo{journal}{Adv. Mater.}} \textbf{\bibinfo{volume}{25}},
  \bibinfo{pages}{4735--4738} (\bibinfo{year}{2013}).

\bibitem{Cancellieri+16nc}
\bibinfo{author}{Cancellieri, C.} \emph{et~al.}
\newblock \bibinfo{title}{Polaronic metal state at the
  $\mathrm{LaAlO_{3}/SrTiO_{3}}$ interface}.
\newblock \emph{\bibinfo{journal}{Nat. Commun.}} \textbf{\bibinfo{volume}{7}},
  \bibinfo{pages}{10386} (\bibinfo{year}{2016}).

\bibitem{Schmitt+12nc}
\bibinfo{author}{Reinle-Schmitt, M.} \emph{et~al.}
\newblock \bibinfo{title}{Tunable conductivity threshold at polar oxide
  interfaces}.
\newblock \emph{\bibinfo{journal}{Nat. Commun.}} \textbf{\bibinfo{volume}{3}},
  \bibinfo{pages}{932} (\bibinfo{year}{2012}).

\bibitem{Pickett+09prl}
\bibinfo{author}{Pentcheva, R.} \& \bibinfo{author}{Pickett, W.~E.}
\newblock \bibinfo{title}{Avoiding the polarization catastrophe in
  $\mathrm{LaAlO_{3}}$ overlayers on $\mathrm{SrTiO_{3}}$(001) through polar
  distortion}.
\newblock \emph{\bibinfo{journal}{Phys. Rev. Lett.}}
  \textbf{\bibinfo{volume}{102}}, \bibinfo{pages}{107602}
  (\bibinfo{year}{2009}).

\bibitem{Stengel+11prl}
\bibinfo{author}{Stengel, M.}
\newblock \bibinfo{title}{First-principles modeling of electrostatically doped
  perovskite systems}.
\newblock \emph{\bibinfo{journal}{Phys. Rev. Lett.}}
  \textbf{\bibinfo{volume}{106}}, \bibinfo{pages}{136803}
  (\bibinfo{year}{2011}).

\bibitem{Cantoni+12am}
\bibinfo{author}{Cantoni, C.} \emph{et~al.}
\newblock \bibinfo{title}{Electron transfer and ionic displacements at the
  origin of the $\mathrm{2D}$ electron gas at the $\mathrm{LAO/STO}$ interface:
  Direct measurements with atomic-column spatial resolution}.
\newblock \emph{\bibinfo{journal}{Adv. Mater.}} \textbf{\bibinfo{volume}{24}},
  \bibinfo{pages}{3952--3957} (\bibinfo{year}{2012}).

\bibitem{Gazquez+17prl}
\bibinfo{author}{Gazquez, J.} \emph{et~al.}
\newblock \bibinfo{title}{Competition between polar and nonpolar lattice
  distortions in oxide quantum wells: New critical thickness at polar
  interfaces}.
\newblock \emph{\bibinfo{journal}{Phys. Rev. Lett.}}
  \textbf{\bibinfo{volume}{119}}, \bibinfo{pages}{106102}
  (\bibinfo{year}{2017}).

\bibitem{Hkim+12ape}
\bibinfo{author}{Kim, H.~J.} \emph{et~al.}
\newblock \bibinfo{title}{High mobility in a stable transparent perovskite
  oxide}.
\newblock \emph{\bibinfo{journal}{Appl. Phys. Express}}
  \textbf{\bibinfo{volume}{5}}, \bibinfo{pages}{061102} (\bibinfo{year}{2012}).

\bibitem{paudel+17prb}
\bibinfo{author}{Paudel, T.~R.} \& \bibinfo{author}{Tsymbal, E.~Y.}
\newblock \bibinfo{title}{Prediction of a mobile two-dimensional electron gas
  at the $\mathrm{LaScO_{3}/BaSnO_{3}}$(001) interface}.
\newblock \emph{\bibinfo{journal}{Phys. Rev. B}} \textbf{\bibinfo{volume}{96}},
  \bibinfo{pages}{245423} (\bibinfo{year}{2017}).

\bibitem{yaqin+16pccp}
\bibinfo{author}{Wang, Y.}, \bibinfo{author}{Tang, W.}, \bibinfo{author}{Cheng,
  J.}, \bibinfo{author}{Nazir, S.} \& \bibinfo{author}{Yang, K.}
\newblock \bibinfo{title}{High-mobility two-dimensional electron gas in
  $\mathrm{SrGeO_{3}}$- and $\mathrm{BaSnO_{3}}$-based perovskite oxide
  heterostructures: an ab initio study}.
\newblock \emph{\bibinfo{journal}{Phys. Chem. Chem. Phys.}}
  \textbf{\bibinfo{volume}{18}}, \bibinfo{pages}{31924--31929}
  (\bibinfo{year}{2016}).

\bibitem{Krish+16apl}
\bibinfo{author}{Krishnaswamy, K.} \emph{et~al.}
\newblock \bibinfo{title}{$\mathrm{BaSnO_{3}}$ as a channel material in
  perovskite oxide heterostructures}.
\newblock \emph{\bibinfo{journal}{Appl. Phys. Lett.}}
  \textbf{\bibinfo{volume}{108}}, \bibinfo{pages}{083501}
  (\bibinfo{year}{2016}).

\bibitem{Useong+15apl}
\bibinfo{author}{Kim, U.} \emph{et~al.}
\newblock \bibinfo{title}{All-perovskite transparent high mobility field effect
  using epitaxial $\mathrm{BaSnO_{3}}$ and $\mathrm{LaInO_{3}}$}.
\newblock \emph{\bibinfo{journal}{APL Mater.}} \textbf{\bibinfo{volume}{3}},
  \bibinfo{pages}{036101} (\bibinfo{year}{2015}).

\bibitem{Lee+17armr}
\bibinfo{author}{Lee, W.-J.} \emph{et~al.}
\newblock \bibinfo{title}{Transparent perovskite barium stannate with high
  electron mobility and thermal stability}.
\newblock \emph{\bibinfo{journal}{Annu. Rev. Mater. Res.}}
  \textbf{\bibinfo{volume}{47}}, \bibinfo{pages}{391--423}
  (\bibinfo{year}{2017}).

\bibitem{Niedermeier+17aps}
\bibinfo{author}{Niedermeier, C.~A.} \emph{et~al.}
\newblock \bibinfo{title}{Electron effective mass and mobility limits in
  degenerate perovskite stannate $\mathrm{BaSnO_{3}}$}.
\newblock \emph{\bibinfo{journal}{Phys. Rev. B}} \textbf{\bibinfo{volume}{95}},
  \bibinfo{pages}{161202} (\bibinfo{year}{2017}).

\bibitem{Niedermeier+16arxiv}
\bibinfo{author}{Niedermeier, C.~A.}, \bibinfo{author}{Kamiya, T.} \&
  \bibinfo{author}{Moram, M.~A.}
\newblock \bibinfo{title}{Polaron coupling constants in $\mathrm{BaSnO}_{3}$}.
\newblock \emph{\bibinfo{journal}{Preprint at
  https://arxiv.org/pdf/1612.01343v1.pdf (2016)}} .

\bibitem{chambers+16apl}
\bibinfo{author}{Chambers, S.~A.}, \bibinfo{author}{Kaspar, T.~C.},
  \bibinfo{author}{Prakash, A.}, \bibinfo{author}{Haugstad, G.} \&
  \bibinfo{author}{Jalan, B.}
\newblock \bibinfo{title}{Band alignment at epitaxial
  $\mathrm{BaSnO_{3}/SrTiO_{3}}$(001) and $\mathrm{BaSnO_{3}/LaAlO_{3}}$(001)
  heterojunctions}.
\newblock \emph{\bibinfo{journal}{Appl. Phys. Lett.}}
  \textbf{\bibinfo{volume}{108}}, \bibinfo{pages}{152104}
  (\bibinfo{year}{2016}).

\bibitem{kim+18apl}
\bibinfo{author}{Kim, Y.}, \bibinfo{author}{Kim, Y.~M.}, \bibinfo{author}{Shin,
  J.} \& \bibinfo{author}{Char, K.}
\newblock \bibinfo{title}{$\mathrm{LaInO_{3}/BaSnO_{3}}$ polar interface on
  $\mathrm{MgO}$ substrates}.
\newblock \emph{\bibinfo{journal}{APL Mater.}} \textbf{\bibinfo{volume}{6}},
  \bibinfo{pages}{096104} (\bibinfo{year}{2018}).

\bibitem{Markurt+19sr}
\bibinfo{author}{Kim, Y.~M.} \emph{et~al.}
\newblock \bibinfo{title}{Interface polarization model for a 2-dimensional
  electron gas at the $\mathrm{BaSnO_{3}/LaInO_{3}}$ interface}.
\newblock \emph{\bibinfo{journal}{Sci. Rep.}} \textbf{\bibinfo{volume}{9}},
  \bibinfo{pages}{16202} (\bibinfo{year}{2019}).

\bibitem{Martina+20prm}
\bibinfo{author}{Zupancic, M.} \emph{et~al.}
\newblock \bibinfo{title}{Role of the interface in controlling the epitaxial
  relationship between orthorhombic $\mathrm{LaInO_{3}}$ and cubic
  $\mathrm{BaSnO_{3}}$}.
\newblock \emph{\bibinfo{journal}{Phys. Rev. Materials}}
  \textbf{\bibinfo{volume}{4}}, \bibinfo{pages}{123605} (\bibinfo{year}{2020}).

\bibitem{Martina+21}
\bibinfo{author}{Zupancic, M.}, \bibinfo{author}{Aggoune, W.},
  \bibinfo{author}{Draxl, C.} \& \bibinfo{author}{Albrecht, M.}
\newblock \bibinfo{title}{Termination at the $\mathrm{BaSnO}_{3}$ surface and
  $\mathrm{LaInO_{3}/BaSnO_{3}}$ interface}.
\newblock \emph{\bibinfo{journal}{In preparation}}  (\bibinfo{year}{2021}).

\bibitem{Campbell+18nm}
\bibinfo{author}{Lee, H.} \emph{et~al.}
\newblock \bibinfo{title}{Direct observation of a two-dimensional hole gas at
  oxide interfaces}.
\newblock \emph{\bibinfo{journal}{Nat. Mater.}} \textbf{\bibinfo{volume}{17}},
  \bibinfo{pages}{231} (\bibinfo{year}{2018}).

\bibitem{Zbigniew+20jcg}
\bibinfo{author}{Galazka, Z.} \emph{et~al.}
\newblock \bibinfo{title}{Melt growth and physical properties of bulk laino3
  single crystals}.
\newblock \emph{\bibinfo{journal}{physica status solidi (a)}}
  \textbf{\bibinfo{volume}{218}}, \bibinfo{pages}{2100016}
  (\bibinfo{year}{2021}).

\bibitem{note-si}
\bibinfo{note}{See Supplemental Material at http:/link....org/ for
  computational details as well as for additional information on structural and
  electronic properties of the pp-type and other selected heterostructures. The
  Born effective charges in pristine systems are computed by
  Ref.~\cite{gula+14jpcm}.}

\bibitem{glazer+72acsb}
\bibinfo{author}{Glazer, A.~M.}
\newblock \bibinfo{title}{The classification of tilted octahedra in
  perovskites}.
\newblock \emph{\bibinfo{journal}{Acta Crystallogr., Sect. B}}
  \textbf{\bibinfo{volume}{28}}, \bibinfo{pages}{3384--3392}
  (\bibinfo{year}{1972}).

\bibitem{Aggoune+BSO}
\bibinfo{author}{Aggoune, W.} \emph{et~al.}
\newblock \bibinfo{title}{Electronic and optical properties of cubic
  $\mathrm{BaSnO}_{3}$: a combined theoretical and experimental study}.
\newblock \emph{\bibinfo{journal}{Preprint at https://arxiv.org/abs/2105.07817
  (2021)}} .

\bibitem{Aggoune+LIO}
\bibinfo{author}{Aggoune, W.} \emph{et~al.}
\newblock \bibinfo{title}{Fingerprints of optical absorption in the perovskite
  $\mathrm{LaInO}_{3}$: Insight from many-body theory and experiment}.
\newblock \emph{\bibinfo{journal}{Phys. Rev. B}}
  \textbf{\bibinfo{volume}{103}}, \bibinfo{pages}{115105}
  (\bibinfo{year}{2021}).

\bibitem{Maznichenko+20pssb}
\bibinfo{author}{Maznichenko, I.~V.}, \bibinfo{author}{Ostanin, S.},
  \bibinfo{author}{Ernst, A.}, \bibinfo{author}{Henk, J.} \&
  \bibinfo{author}{Mertig, I.}
\newblock \bibinfo{title}{Formation and tuning of $\mathrm{2D}$ electron gas in
  perovskite heterostructures}.
\newblock \emph{\bibinfo{journal}{Phys. Status Solidi B}}
  \textbf{\bibinfo{volume}{257}}, \bibinfo{pages}{1900540}
  (\bibinfo{year}{2020}).

\bibitem{Millis+10prb}
\bibinfo{author}{Millis, A.~J.} \& \bibinfo{author}{Schlom, D.~G.}
\newblock \bibinfo{title}{Electron-hole liquids in transition-metal oxide
  heterostructures}.
\newblock \emph{\bibinfo{journal}{Phys. Rev. B}} \textbf{\bibinfo{volume}{82}},
  \bibinfo{pages}{073101} (\bibinfo{year}{2010}).

\bibitem{Eisenstein+04n}
\bibinfo{author}{Eisenstein, J.} \& \bibinfo{author}{MacDonald, A.}
\newblock \bibinfo{title}{Bose–einstein condensation of excitons in bilayer
  electron systems}.
\newblock \emph{\bibinfo{journal}{Nature}} \textbf{\bibinfo{volume}{432}},
  \bibinfo{pages}{691–694} (\bibinfo{year}{2004}).

\bibitem{berryPhase+93prbr}
\bibinfo{author}{King-Smith, R.~D.} \& \bibinfo{author}{Vanderbilt, D.}
\newblock \bibinfo{title}{Theory of polarization of crystalline solids}.
\newblock \emph{\bibinfo{journal}{Phys. Rev. B}} \textbf{\bibinfo{volume}{47}},
  \bibinfo{pages}{1651--1654} (\bibinfo{year}{1993}).

\bibitem{Resta+93prl}
\bibinfo{author}{Resta, R.}, \bibinfo{author}{Posternak, M.} \&
  \bibinfo{author}{Baldereschi, A.}
\newblock \bibinfo{title}{Towards a quantum theory of polarization in
  ferroelectrics: The case of $\mathrm{KNbO_{3}}$}.
\newblock \emph{\bibinfo{journal}{Phys. Rev. Lett.}}
  \textbf{\bibinfo{volume}{70}}, \bibinfo{pages}{1010--1013}
  (\bibinfo{year}{1993}).

\bibitem{jang+17jap}
\bibinfo{author}{Jang, D.~H.} \emph{et~al.}
\newblock \bibinfo{title}{Single crystal growth and optical properties of a
  transparent perovskite oxide $\mathrm{LaInO_{3}}$}.
\newblock \emph{\bibinfo{journal}{J. Appl. Phys.}}
  \textbf{\bibinfo{volume}{121}}, \bibinfo{pages}{125109}
  (\bibinfo{year}{2017}).

\bibitem{Zhou+19prr}
\bibinfo{author}{Zhou, J.-J.} \& \bibinfo{author}{Bernardi, M.}
\newblock \bibinfo{title}{Predicting charge transport in the presence of
  polarons: The beyond-quasiparticle regime in $\mathrm{SrTiO_{3}}$}.
\newblock \emph{\bibinfo{journal}{Phys. Rev. Research}}
  \textbf{\bibinfo{volume}{1}}, \bibinfo{pages}{033138} (\bibinfo{year}{2019}).

\bibitem{krish+17prb}
\bibinfo{author}{Krishnaswamy, K.}, \bibinfo{author}{Himmetoglu, B.},
  \bibinfo{author}{Kang, Y.}, \bibinfo{author}{Janotti, A.} \&
  \bibinfo{author}{Van~de Walle, C.~G.}
\newblock \bibinfo{title}{First-principles analysis of electron transport in
  $\mathrm{BaSnO_{3}}$}.
\newblock \emph{\bibinfo{journal}{Phys. Rev. B}} \textbf{\bibinfo{volume}{95}},
  \bibinfo{pages}{205202} (\bibinfo{year}{2017}).

\bibitem{dima+19cm}
\bibinfo{author}{Nabok, D.}, \bibinfo{author}{Höffling, B.} \&
  \bibinfo{author}{Draxl, C.}
\newblock \bibinfo{title}{Energy-level alignment at organic/inorganic
  interfaces from first principles: Example of poly(para-phenylene)/rock-salt
  $\mathrm{ZnO}$(100)}.
\newblock \emph{\bibinfo{journal}{Chem. Mater.}} \textbf{\bibinfo{volume}{31}},
  \bibinfo{pages}{7143--7150} (\bibinfo{year}{2019}).

\bibitem{galazka+16jpcm}
\bibinfo{author}{Galazka, Z.} \emph{et~al.}
\newblock \bibinfo{title}{Melt growth and properties of bulk
  $\mathrm{BaSnO}_{3}$ single crystals}.
\newblock \emph{\bibinfo{journal}{J. Phys. Condens. Matter}}
  \textbf{\bibinfo{volume}{29}}, \bibinfo{pages}{075701}
  (\bibinfo{year}{2016}).

\bibitem{PBEsol+08prl}
\bibinfo{author}{Perdew, J.~P.} \emph{et~al.}
\newblock \bibinfo{title}{Restoring the density-gradient expansion for exchange
  in solids and surfaces}.
\newblock \emph{\bibinfo{journal}{Phys. Rev. Lett.}}
  \textbf{\bibinfo{volume}{100}}, \bibinfo{pages}{136406}
  (\bibinfo{year}{2008}).

\bibitem{FHI-aims}
\bibinfo{author}{Blum, V.} \emph{et~al.}
\newblock \bibinfo{title}{Ab initio molecular simulations with numeric
  atom-centered orbitals}.
\newblock \emph{\bibinfo{journal}{Comput. Phys. Commun}}
  \textbf{\bibinfo{volume}{180}}, \bibinfo{pages}{2175 -- 2196}
  (\bibinfo{year}{2009}).

\bibitem{momm-izum11jacr}
\bibinfo{author}{Momma, K.} \& \bibinfo{author}{Izumi, F.}
\newblock \bibinfo{title}{{{\it VESTA3} for three-dimensional visualization of
  crystal, volumetric and morphology data}}.
\newblock \emph{\bibinfo{journal}{J.~Appl.~Cryst.~}}
  \textbf{\bibinfo{volume}{44}}, \bibinfo{pages}{1272--1276}
  (\bibinfo{year}{2011}).

\bibitem{drax-sche19jpm}
\bibinfo{author}{Draxl, C.} \& \bibinfo{author}{Scheffler, M.}
\newblock \bibinfo{title}{The nomad laboratory: from data sharing to artificial
  intelligence}.
\newblock \emph{\bibinfo{journal}{J.~Phys.~Mater.~}}
  \textbf{\bibinfo{volume}{2}}, \bibinfo{pages}{036001} (\bibinfo{year}{2019}).

\bibitem{gula+14jpcm}
\bibinfo{author}{Gulans, A.} \emph{et~al.}
\newblock \bibinfo{title}{exciting: a full-potential all-electron package
  implementing density-functional theory and many-body perturbation theory}.
\newblock \emph{\bibinfo{journal}{J.~Phys.~Condens.~Matter.~}}
  \textbf{\bibinfo{volume}{26}}, \bibinfo{pages}{363202}
  (\bibinfo{year}{2014}).

\end{thebibliography}

\begin{thebibliography}{10}
\expandafter\ifx\csname url\endcsname\relax
  \def\url#1{\texttt{#1}}\fi
\expandafter\ifx\csname urlprefix\endcsname\relax\def\urlprefix{URL }\fi
\providecommand{\bibinfo}[2]{#2}
\providecommand{\eprint}[2][]{\url{#2}}

\bibitem{Markurt+19sr}
\bibinfo{author}{Kim, Y.~M.} \emph{et~al.}
\newblock \bibinfo{title}{Interface polarization model for a 2-dimensional
  electron gas at the $\mathrm{BaSnO_{3}/LaInO_{3}}$ interface}.
\newblock \emph{\bibinfo{journal}{Sci. Rep.}} \textbf{\bibinfo{volume}{9}},
  \bibinfo{pages}{16202} (\bibinfo{year}{2019}).

\bibitem{Pickett+09prl}
\bibinfo{author}{Pentcheva, R.} \& \bibinfo{author}{Pickett, W.~E.}
\newblock \bibinfo{title}{Avoiding the polarization catastrophe in
  $\mathrm{LaAlO_{3}}$ overlayers on $\mathrm{SrTiO_{3}}$(001) through polar
  distortion}.
\newblock \emph{\bibinfo{journal}{Phys. Rev. Lett.}}
  \textbf{\bibinfo{volume}{102}}, \bibinfo{pages}{107602}
  (\bibinfo{year}{2009}).

\bibitem{berryPhase+93prbr}
\bibinfo{author}{King-Smith, R.~D.} \& \bibinfo{author}{Vanderbilt, D.}
\newblock \bibinfo{title}{Theory of polarization of crystalline solids}.
\newblock \emph{\bibinfo{journal}{Phys. Rev. B}} \textbf{\bibinfo{volume}{47}},
  \bibinfo{pages}{1651--1654} (\bibinfo{year}{1993}).

\bibitem{gula+14jpcm}
\bibinfo{author}{Gulans, A.} \emph{et~al.}
\newblock \bibinfo{title}{exciting: a full-potential all-electron package
  implementing density-functional theory and many-body perturbation theory}.
\newblock \emph{\bibinfo{journal}{J.~Phys.~Condens.~Matter.~}}
  \textbf{\bibinfo{volume}{26}}, \bibinfo{pages}{363202}
  (\bibinfo{year}{2014}).

\bibitem{Gazquez+17prl}
\bibinfo{author}{Gazquez, J.} \emph{et~al.}
\newblock \bibinfo{title}{Competition between polar and nonpolar lattice
  distortions in oxide quantum wells: New critical thickness at polar
  interfaces}.
\newblock \emph{\bibinfo{journal}{Phys. Rev. Lett.}}
  \textbf{\bibinfo{volume}{119}}, \bibinfo{pages}{106102}
  (\bibinfo{year}{2017}).

\bibitem{Resta+93prl}
\bibinfo{author}{Resta, R.}, \bibinfo{author}{Posternak, M.} \&
  \bibinfo{author}{Baldereschi, A.}
\newblock \bibinfo{title}{Towards a quantum theory of polarization in
  ferroelectrics: The case of $\mathrm{KNbO_{3}}$}.
\newblock \emph{\bibinfo{journal}{Phys. Rev. Lett.}}
  \textbf{\bibinfo{volume}{70}}, \bibinfo{pages}{1010--1013}
  (\bibinfo{year}{1993}).

\bibitem{Aggoune+LIO}
\bibinfo{author}{Aggoune, W.} \emph{et~al.}
\newblock \bibinfo{title}{Fingerprints of optical absorption in the perovskite
  $\mathrm{LaInO}_{3}$: Insight from many-body theory and experiment}.
\newblock \emph{\bibinfo{journal}{Phys. Rev. B}}
  \textbf{\bibinfo{volume}{103}}, \bibinfo{pages}{115105}
  (\bibinfo{year}{2021}).

\bibitem{Martina+21}
\bibinfo{author}{Zupancic, M.}, \bibinfo{author}{Aggoune, W.},
  \bibinfo{author}{Draxl, C.} \& \bibinfo{author}{Albrecht, M.}
\newblock \bibinfo{title}{Termination at the $\mathrm{BaSnO}_{3}$ surface and
  $\mathrm{LaInO_{3}/BaSnO_{3}}$ interface}.
\newblock \emph{\bibinfo{journal}{In preparation}}  (\bibinfo{year}{2021}).

\bibitem{Aggoune+BSO}
\bibinfo{author}{Aggoune, W.} \emph{et~al.}
\newblock \bibinfo{title}{Electronic and optical properties of cubic
  $\mathrm{BaSnO}_{3}$: a combined theoretical and experimental study}.
\newblock \emph{\bibinfo{journal}{Preprint at https://arxiv.org/abs/2105.07817
  (2021)}} .

\bibitem{dima+19cm}
\bibinfo{author}{Nabok, D.}, \bibinfo{author}{Höffling, B.} \&
  \bibinfo{author}{Draxl, C.}
\newblock \bibinfo{title}{Energy-level alignment at organic/inorganic
  interfaces from first principles: Example of poly(para-phenylene)/rock-salt
  $\mathrm{ZnO}$(100)}.
\newblock \emph{\bibinfo{journal}{Chem. Mater.}} \textbf{\bibinfo{volume}{31}},
  \bibinfo{pages}{7143--7150} (\bibinfo{year}{2019}).

\end{thebibliography}
\end{document}